\documentclass[12pt,twoside]{article}

\def\TheAuthor{Norbert Schuch}                              
\def\TheTitle{Condensed Matter Applications of Entanglement Theory\footnotemark}                        
\def\TheHeading{Condensed Matter Applications of Entanglement Theory}
\def\TheInstitute{Peter Gr\"unberg Institut}                                         
\def\TheAddress{Forschungszentrum J\"ulich GmbH}                           
\def\TheInstitute{Institute for Quantum Information}  
\def\TheAddress{RWTH Aachen, 52056 Aachen, Germany}         
\def\ThePart{C}                                                                                  
\def\TheChapter{4}                                                                            

\usepackage{ferienschule_12}                                  
\usepackage{color}
\usepackage{cite}

\newcommand{\ket}[1]{\vert#1\rangle}
\newcommand{\bra}[1]{\langle #1\vert}
\newcommand{\mc}[1]{\mathcal{#1}}
\newcommand{\tr}{\mathrm{tr}}
\def\openone{\leavevmode\hbox{\small1\kern-3.8pt\normalsize1}}

\begin{document}

\MakeTitel           

\setcounter{tocdepth}{2}
\tableofcontents     

\footnotetext{Lecture Notes of the $44^{{\rm th}}$ IFF Spring School
``Quantum Information Processing'', edited by D.\ DiVincenzo
(Forschungszentrum J{\"{u}}lich, 2013).}

\newpage

\section{Introduction}

The aim of this lecture is to show how the tools of quantum information
theory and in particular entanglement theory can help us to understand the
physics of condensed matter systems and other correlated quantum
many-body systems. In a certain sense, many-body systems are all around in
quantum information: The defining feature of these systems is that they
consist of many subsystems of the same dimension, equipped with a
natural tensor product structure\footnote{We ignore fermionic systems for
the moment, but as it turns out, many of the results described in this
lecture also hold for fermionic systems, see Sec.~\ref{ssec:fermions}.}
$(\mathbb C^d)^{\otimes N}$, and such structures appear in quantum
information as quantum registers in quantum computers, as parties in
multipartite protocols, and so on.

Most condensed matter systems exhibit only weak quantum correlations (this
is, entanglement) between the individual subsystems. These systems are well
described by a mean field ansatz (i.e., a product state), and their
behavior can be 
understood using Landau theory.  However, some of the most exciting
phenomena discovered in condensed matter physics in the last decades, such
as the fraction quantum Hall effect or high-temperature superconductivity,
are based on systems with strong interactions in which entanglement
plays an essential role.  Given the refined understanding of
entanglement which has been developed in the field of quantum information
theory, it is thus natural to apply quantum information concepts, and in
particular the theory of entanglement, towards an improved understanding of quantum
many-body systems.  Indeed, an active field of research has grown during
the last decade at the interface of quantum information theory and quantum
many-body physics, and the aim of this lecture is to give an introduction
to this area. 

In this lecture, we will highlight two complementary topics at the interface between
condensed matter and quantum information: In the first part
(Sec.~\ref{sec:mps},~\ref{sec:2d-and-beyond}), we will show how we
can use insights on the entanglement structure of condensed matter systems
to develop powerful methods to both numerically simulate and analytically
characterize correlated many-body systems; and in the second part
(Sec.~\ref{sec:cplx}) we will show how the field of quantum complexity
theory allows us to better understand the limitations to our (numerical)
understanding of those systems.

\subsection{\label{ssec:arealaw}
Entanglement structure of ground states: The area law}

For clarity of the presentation, we will in the following restrict our
attention to quantum spin systems on a lattice (such as a line or a square
lattice in 2D), with a corresponding Hilbert space $(\mathbb C^d)^{\otimes
N}$ (where each spin has $d$ levels, and the lattice has $N$ sites);
generalizations to fermionic systems and beyond lattices will be discussed
later.  Also, we will for the moment focus on ground state problems, i.e.,
given some Hamiltonian $H$ acting or our spin system, we will ask about
properties of its ground state $\ket\Psi$.  The approach we pursue will be
variational -- we will try to obtain a family of states which gives a good
approximation of the ground state, for which quantities of interest can be
evaluated efficiently, and where the best approximation to the ground
state can be found efficiently.  For instance, mean-field theory is a
variational theory based on the class of product states (for spin systems)
or Slater determinants (for electronic systems).

So which states should we use for our variational description of quantum
spin systems? Of course, one could simply try to parametrize the ground state as
\begin{equation}
    \label{eq:expand-psi}
\ket\Psi=\sum_{i_1,\dots,i_N} c_{i_1\dots i_N}\ket{i_1,\dots,i_N}\ ,
\end{equation}
and use the $c_{i_1\dots i_N}$ as variational parameters.  Unfortunately,
the number of parameters $c_{i_1\dots i_N}$ grows exponentially with $N$,
making it impossible to have an efficient description of $\ket\Psi$ for
growing system sizes.
On the other hand, we know that efficient descriptions exist for
physical Hamiltonians: Since $H=\sum_{i} h_i$ is a sum of few-body terms
(even if we don't restrict to lattice systems), a polynomial number $N^k$
of parameters (with $k$ the bodiness of the interaction) allows to
specify $H$, and thus its ground state.  This is, while a general $N$-body
quantum state can occupy an exponentially large Hilbert space, all
physical states live in a very small ``corner'' of this space.  The
difficulty, of course, is to find an efficient parametrization which
captures the states in this corner of
Hilbert space, while at the same time allowing for efficient simulation
methods.

\begin{figure}
 \centering
 \includegraphics[height=2cm]{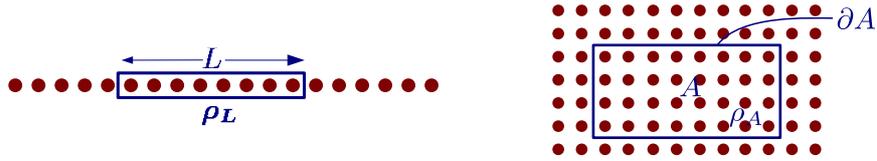}
 \caption{\label{fig:arealaw}
    Area law: The entropy of the reduced states of a block $A$ scales like
    the length of its boundary $\partial A$; in one dimension, this
    implies that the entropy is bounded by a constant.
}
\end{figure}

In order to have a guideline for constructing an ansatz class,  we choose
to look
at the entanglement properties of ground states of interacting quantum
systems. To this end, we consider a ground state $\ket\Psi$ on a lattice
and cut a contiguous region of length $L$ (in one dimension) or an area $A$
(in two dimensions), cf.~Fig.~\ref{fig:arealaw}.
It is a well-known result from entanglement
theory~\cite{bennett:purestate-ent} that the von Neumann entropy of the
reduced density matrix $\rho_A$ of region $A$,
\[
S(\rho_A)=-\rho_A\log\,\rho_A\ ,
\]
quantifies the entanglement between region $A$ and the rest of the system.
For a random quantum state, we expect this entanglement to be almost
maximal, i.e., on the order of $|A|\log d$ (where $|A|$ is the number of
spins in region $A$). Yet, if we study the behavior of $S(\rho_A)$ for ground
states of local Hamiltonians, it is found that $S(\rho_A)$ essentially scales
like the \emph{boundary} of region $A$, $S(\rho_A)\propto |\partial A|$, with
possible corrections for gapless Hamiltonians which are at most logarithmic
in the volume, $S(\rho_A)\propto
|\partial A|\log|A|$.  This behavior is known as the \emph{area
law} 
\index{area law}
for the entanglement entropy and has been observed throughout for
ground states of local Hamiltonians~(see, e.g.,
Ref.~\cite{eisert:arealaw-review} for
a review); for gapped Hamiltonians in one dimension, this result has been
recently proven rigorously~\cite{hastings:arealaw}.

\section{One-dimensional systems: Matrix Product States
\label{sec:mps}}

\subsection{Matrix Product States\label{ssec:mps}}

\subsubsection{Construction and Matrix Product notation}

\begin{figure}
 \centering
 \includegraphics[height=2cm]{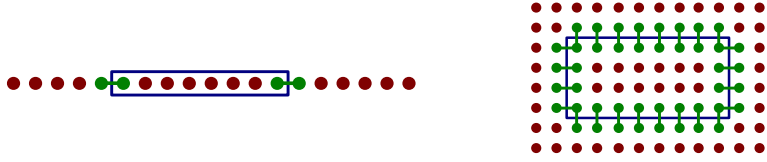}
 \caption{\label{fig:arealaw-2}
    The area law suggests that the entanglement between two regions is
    located around the boundary.
}
\end{figure}

\begin{figure}[t]
 \centering
 \includegraphics[width=\textwidth]{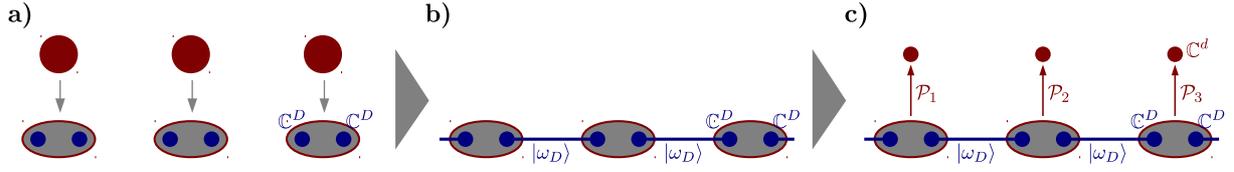}
 \caption{\label{fig:mps-constr}
    Construction of MPS: \textbf{a)} Each site is composed of two virtual
    subsystems. \textbf{b)} The virtual subsystems are placed in maximally
    entangled states. \textbf{c)} Linear maps $\mc P_s$ are applied which
    map the two virtual systems to the physical system.
}
\end{figure}

Since the entropy $S(\rho_A)$ quantifies the entanglement between region
$A$ and its complement, the fact that $S(\rho_A)$ scales like the
\emph{boundary} of $\rho_A$ suggests that the entanglement between region
$A$ and the rest of the system is essentially located around the boundary
between the two regions, as illustrated in Fig.~\ref{fig:arealaw-2}.
We will now construct an ansatz for many-body quantum systems, starting
from the insight that the entanglement is concentrated around the boundary
of regions; for the moment, we will focus on one-dimensional
systems.  Clearly, since we want to have this property for any
partitioning of the lattice, we cannot just place entangled pairs as in
Fig.~\ref{fig:arealaw-2}, but we have to choose a more subtle strategy.
To this end, we consider the system at each site as being composed of two
``virtual'' subsystems of dimension $D$ each, as illustrated in
Fig.~\ref{fig:mps-constr}a. Then, each of the two subsystems is placed in
a maximally entangled state 
\[
\ket{\omega_D}=\sum_{i=1}^D\ket{i,i}
\] 
with the corresponding subsystems of the adjacent sites, as shown in 
Fig.~\ref{fig:mps-constr}b. The maximally entangled states are called
``bonds'', with $D$ the ``bond dimension''.  This construction already
satisfies the area law: For any region we cut, there are exactly two
maximally entangled states crossing the cuts, bounding the entanglement by
$2\log D$.  Finally, we apply linear maps $\mc P_s:\mathbb C^D\otimes
\mathbb C^D\rightarrow
\mathbb C^d$ at each site $s$, which creates a description of a state on a
chain of $d$-level systems, cf.~Fig.~\ref{fig:mps-constr}c.  
(Note that the rank of the reduced density operator of any region cannot
be increased by applying the linear maps $\mc P_s$.) 
The construction can be carried out either
with periodic boundary conditions, or with open boundary conditions by
omitting the outermost virtual subsystems at the end of the chain.  The
total state of the chain can 
be written as
\begin{equation}
    \label{eq:mps-as-peps}
\ket{\Psi}=(\mc P_1\otimes\cdots \otimes \mc P_N) \ket{\omega_{D}}^{\otimes N}\
,
\end{equation}
where the maps $\mc P_s$ act on the maximally entangled states as
illustrated in 
Fig.~\ref{fig:mps-constr}c.

This class of states can be rewritten as follows: For each site $s$, define a
three-index tensor $A^{[s]}_{i,\alpha\beta}$, $i=1,\dots,d$,
$\alpha,\beta=1,\dots,D$, where the $A^{[s]}_i$ can be interpreted as
$D\times D$ matrices, such that 
\begin{equation}
\mc P_s = \sum_{i,\alpha,\beta} A^{[s]}_{i,\alpha\beta}
    \ket{i}\bra{\alpha,\beta}\ .
\end{equation}
Then, the state (\ref{eq:mps-as-peps}) can be
rewritten as
\begin{equation}
    \label{eq:mps-def}
\ket\Psi = \sum_{i_1,\dots,i_N} \tr\big[A^{[1]}_{i_1}
    A^{[2]}_{i_2} \cdots A^{[N]}_{i_N}\big]\;\ket{i_1,\dots,i_N}\ ,
\end{equation}
i.e., the coefficient $c_{i_1\dots i_N}$ in (\ref{eq:expand-psi}) can be
expressed as a product of matrices.\footnote{
The equivalence of (\ref{eq:mps-as-peps}) and (\ref{eq:mps-def}) can be
proven straightforwardly by noting that for two maps $\mc P_1$ and $\mc
P_2$, and the bond $\ket{\omega_D}$ between them, it holds that
\[
\mc P_1\otimes \mc P_2 \ket{\omega_D} = 
\sum_{i_1,i_2,\alpha,\beta} 
	     (A^{[1]}_{i_1} A^{[2]}_{i_2})_{\alpha\beta}
	    \ket{i_1,i_2}\bra{\alpha,\beta}\ ,
\]
and iterating this argument through the chain.
}  
For this reason, these states are called \emph{Matrix Product States}
(MPS). \index{Matrix Product States (MPS)}  For systems with open
boundary conditions, the matrices $A^{[1]}_{i_1}$ and $A^{[N]}_{i_N}$ 
are $1\times D$ and $D\times 1$ matrices, respectively, so that the trace
can be omitted.  More generally, $D$ can be chosen differently across each
link.

\subsubsection{Tensor network notation}

The defining equation (\ref{eq:mps-def}) for Matrix Product States is a
special case of a so-called \emph{tensor network}. \index{tensor network}
Generally, tensor networks are given by a number of tensors
$A_{i_1,i_2,\dots,i_K}$, $B_{i_1,i_2,\dots,i_K}$, etc., where each tensor
usually only depends on a few of the indices. Then, one takes the product
of the tensors and sums over a subset of the indices,
\[
c_{i_1\dots i_k} = \sum_{i_{k+1},\dots, i_K}
A_{i_1,i_2,\dots,i_K}B_{i_1,i_2,\dots,i_K}\cdots\ .
\]
For instance, in (\ref{eq:mps-def}) the tensors are the $A^{[s]}\equiv
A^{[s]}_{i,\alpha\beta}$, and we sum over the virtual indices
$\alpha,\beta,\dots$, yielding
\[
c_{i_1\dots i_N}
=
\sum_{\alpha,\beta,\gamma,\dots,\zeta}
    A^{[1]}_{i_1,\alpha\beta} A^{[2]}_{i_2,\beta\gamma}\cdots A^{[N]}_{i_N,\zeta\alpha}
=
\tr\big[A^{[1]}_{i_1}
    A^{[2]}_{i_2} \cdots A^{[N]}_{i_N}\big]\ .
\]
Tensor networks are most conveniently expressed in a graphical language.
Each tensor is denoted by a box with ``legs'' attached to it, where each
leg corresponds to an index -- a three-index tensor $A^{[s]}\equiv
A^{[s]}_{i,\alpha\beta}$ is then
depicted as 
\[
\raisebox{-0.2cm}{
\includegraphics[scale=0.55]{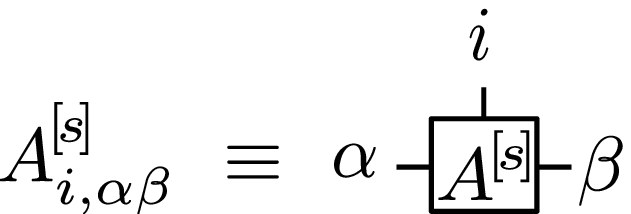}}\quad.
\]
 Summing over a joint index is
denoted by connecting the corresponding legs, e.g., 
\[
\raisebox{-0.4cm}{
\includegraphics[scale=0.55]{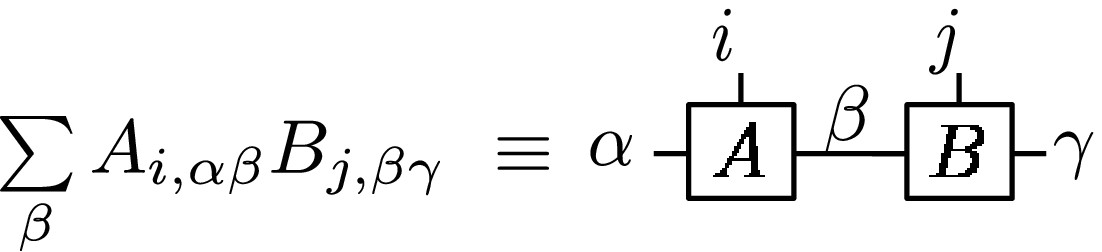}}\quad.
\]
In this language, the expansion coefficient $c_{i_1\dots i_N}$
[Eq.~(\ref{eq:expand-psi})] of an MPS (which we will further on use
interchangably with the state itself) is written
\begin{equation}
\label{eq:mps-as-tn}
\raisebox{-0.5cm}{
\includegraphics[scale=0.55]{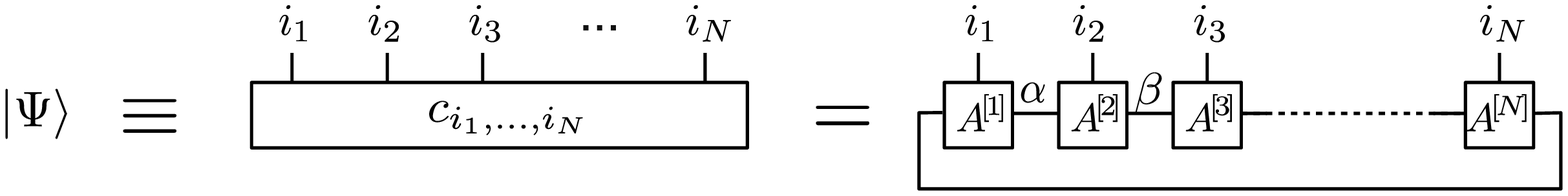}}\quad.
\end{equation}
We will make heavy use of this graphical
language for tensor networks in the following.

\subsubsection{Approximating states as Matrix Product States
\label{sssec:mps-approximability}}

As it turns out, MPS are very well suited to describe ground states of
one-dimensional quantum systems.  On the one hand, we have seen that by
construction, these states all satisfy the area law.   On the other hand,
it can be shown that all states which satisfies an area law, such as ground
states of gapped Hamiltonians~\cite{hastings:arealaw}, as well as states
for which the entanglement of a block grows slowly (such as for critical
1D systems), can be well approximated by an
MPS~\cite{verstraete:faithfully,hastings:arealaw}: Given a state
$\ket{\Phi}$ on a chain of length $N$ for which the entropy of any block
of length $L$ is bounded by $S_\mathrm{max}$,  $S(\rho_L)\le
S_\mathrm{max}$, there exists an MPS $\ket{\Psi_D}$ which approximates
$\ket{\Phi}$ up to error\footnote{
Stricly speaking, this bound only follows from an area law for the
\emph{R\'enyi entropy} 
\[
S_\alpha = \frac{\log \tr[\rho^\alpha]}{1-\alpha}
\]
for $\alpha<1$, with $c$ in (\ref{eq:mps-approx-scaling}) depending on
$\alpha$~\cite{verstraete:faithfully}, which also holds for gapped
Hamiltonians~\cite{hastings:arealaw}.  The proof uses the fact that a bound
on the area law implies a fast decay of the Schmidt coefficients (i.e.,
the eigenvalues of the reduced density operator), and thus, one can
construct an MPS by sequentially doing Schmidt decompositions of the state
and discarding all but the largest $D$ Schmidt
coefficients~\cite{vidal:timeevol,verstraete:faithfully}.
}
\begin{equation}\label{eq:mps-approx-scaling}
|\ket\Phi-\ket{\Psi_D}| =: \epsilon\,
\le 
\mathrm{const} \times
\frac{N  e^{c S_\mathrm{max}}}{
D^{c}}\ .
\end{equation}
Note that even if $S_\mathrm{max}$ grows logarithmically with $N$, the
numerator is still a polynomial in $N$.  This is, in order to achieve a
given accuracy $\epsilon$, we need to choose a bond dimension $D$ which
scales polynomially in $N$ and $1/\epsilon$, and thus, the total number of
parameters (and, as we will see later, also the computation time) scales
polynomially as long as the desired accuracy is at most
$1/\mathrm{poly}(N)$.

\subsubsection{States with exact Matrix Product form}

Beyond the fact that ground states of gapped Hamiltonians can be
efficiently approximated by MPS, there is a large class of states which
can be expressed exactly as MPS with a small bond dimension $D$.  First of
all, any product state is trivially an MPS with $D=1$.  The GHZ state
\[
\ket{\mathrm{GHZ}} = \ket{0\cdots 0}+\ket{1\cdots 1}
\]
is an MPS with $D=2$, with a translational invariant MPS representation
$A^{[s]}\equiv A\ \forall\,s$, with
$A_0=\ket{0}\bra{0}\equiv\left(\begin{smallmatrix}1&0\\0&0\end{smallmatrix}\right)$,
and 
$A_1=\ket{1}\bra{1}\equiv\left(\begin{smallmatrix}0&0\\0&1\end{smallmatrix}\right)$.
The W state
\[
\ket{\mathrm{W}}=\ket{100\cdots00}+\ket{010\cdots00}+\dots+\ket{000\cdots01}
\]
is an MPS with $D=2$, and $A^{[s]}=A$ for $2\le s\le N-1$, 
$A^{[1]}=\bra{0}A$, and $A^{[N]}=A\ket{1}$, where $A_0=\openone$ and
$A_1=\ket{0}\bra{1}$. Other states which have an MPS representation with
small $D$ are the cluster state used in measurement-based
computation~\cite{raussendorf:cluster-short,verstraete:mbc-peps} (with
$D=2$) or the 1D resonating valence bond state which appears as the ground
state of the so-called Majumdar-Ghosh
model~\cite{majumdar:majumdar-ghosh-model,verstraete:comp-power-of-peps}
(with $D=3$), and the AKLT state which we will get to know in
Section~\ref{sssec:aklt-and-parent}.

\subsection{Variational calculation with MPS: The DMRG method}

As we have discussed at the end of
Section~\ref{sssec:mps-approximability},  MPS approximate ground states of
local Hamiltonians efficiently, as the effort needed for a good
approximation scales only polynomially in the length of the chain and the
desired accuracy. Thus, it seems appealing to use the class of MPS as a
variational ansatz to simulate the properties of quantum many-body
systems.  However, to this end it is not sufficient to have an efficient
description of relevant states -- after all, the Hamiltonian itself forms
an efficient description of its ground state, but it is hard to extract
information from it!  Rather, a good variational class also requires that
we can efficiently extract quantities of interest such as energies,
correlation functions, and the like, and that there is an efficient way to
\emph{find} the ground state (i.e., minimize the energy within the
variational class of states) in the first place.

\subsubsection{
\label{ssec:compute-expval}
Evaluating expectation values for MPS}

Let us start by discussing how to compute the expectation value of a local
operator $h$ (such as a term in the Hamiltonian) for an MPS. To this end,
note that 
\[
\bra\Psi h \ket\Psi = 
\sum_{{i_1,\dots,i_N}\atop{j_1,\dots,j_N}} 
c_{i_1\dots i_N}c^*_{j_1\dots j_N} 
\delta_{i_1,j_1}\cdots\delta_{i_{k-1},j_{k-1}}
h_{i_{k}i_{k+1}}^{j_kj_{k+1}}
\delta_{i_{k+2},j_{k+2}}\cdots\delta_{i_N,j_N}
\]
where 
\[
h = \sum_{ {i_k,i_{k+1}}\atop{j_{k},j_{k+1}}}
    h_{i_{k}i_{k+1}}^{j_kj_{k+1}}
    \ket{i_{k},i_{k+1}}\bra{j_k,j_{k+1}}
\]
 acts on sites $k$ and $k+1$.
Using the graphical tensor network notation, this can be written as
\vspace*{+0.1cm}
\begin{equation}
    \label{eq:exp-value-tn}
\raisebox{-.9cm}{\includegraphics[scale=0.55]{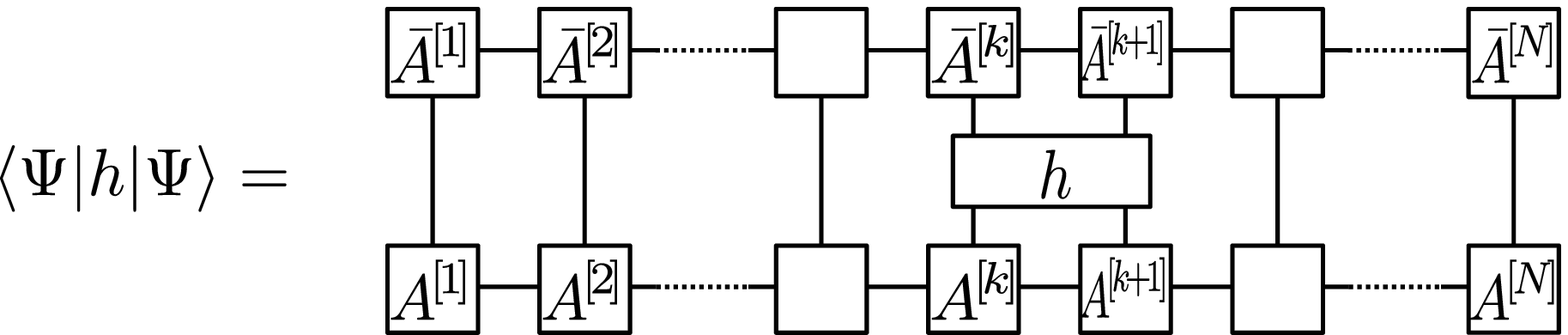}}\qquad .
\vspace*{+0.1cm}
\end{equation}
In order to evaluate this quantity, we have to contract the whole diagram
(\ref{eq:exp-value-tn}). In principle, contracting arbitrary tensor
networks can become an extremely hard problem (strictly speaking,
\textsf{PP}-hard~\cite{schuch:cplx-of-PEPS}), as in some cases it essentially
requires to determine exponentially big tensors (e.g., we might first have
to compute $c_{i_1\dots i_N}$ from the tensor network and from it
determine the expectation value). Fortunately, it turns out that the
tensor network of Eq.~(\ref{eq:exp-value-tn}) can be contracted
efficiently, i.e., with an effort polynomial in $D$ and $N$. To this end,
let us start from the very left of the tensor network in
Eq.~(\ref{eq:exp-value-tn}) and block the leftmost column (tensors
$A^{[1]}$ and $\bar A^{[1]}$). Contracting the internal index, this gives
a two-index tensor
\[
\mathbb L^{\alpha\alpha'} = 
    \sum_i A^{[1]}_{i\alpha} \bar A^{[1]}_{i\alpha'}\ ,
\]
which we interpret as a (bra) \emph{vector} with a ``double index''
$\alpha\alpha'$ of dimension $D^2$. Graphically, this can be denoted as 
\[
\raisebox{-0.9cm}{
\includegraphics[scale=0.55]{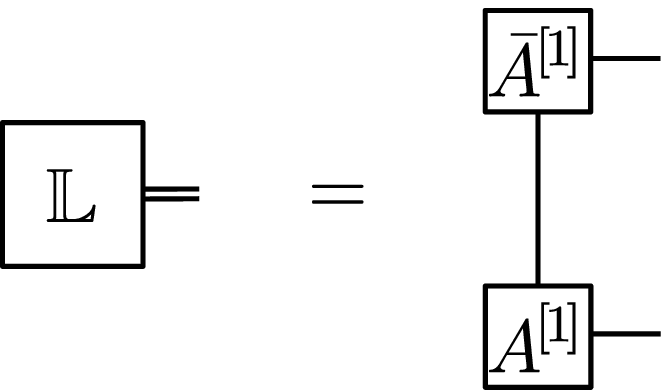}}\quad,
\]
where we use a doubled line to denote the ``doubled'' index of dimension
$D^2$.
We can now continue this way, and define operators (called \emph{transfer
operators})
\index{transfer operator}
\[
(\mathbb E^{[s]})_{\alpha\alpha'}^{\beta\beta'} = 
\sum_i A^{[s]}_{i,\alpha\beta} \bar{A}^{[s]}_{i,\alpha'\beta'}
\]
which we interpret as mapping the double index $\alpha\alpha'$ 
to $\beta\beta'$, and graphically write as
\begin{equation}
\label{eq:top-def-sitedep}
\raisebox{-0.9cm}{
\includegraphics[scale=0.55]{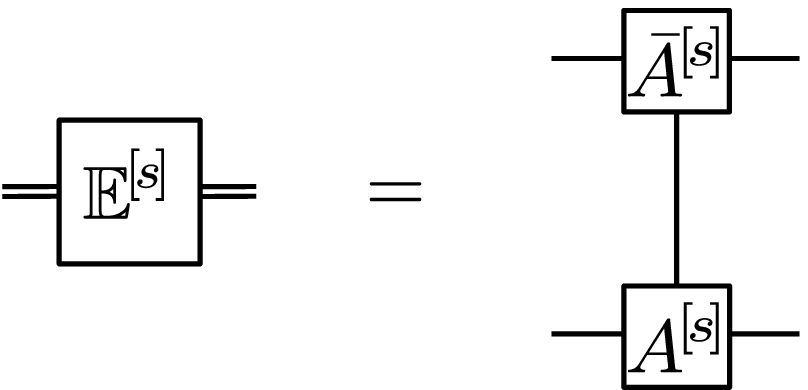}}\quad.
\end{equation}
Similarly, we define operators
\begin{equation}\label{eq:def-top-with-h}
\raisebox{-0.9cm}{
\includegraphics[scale=0.55]{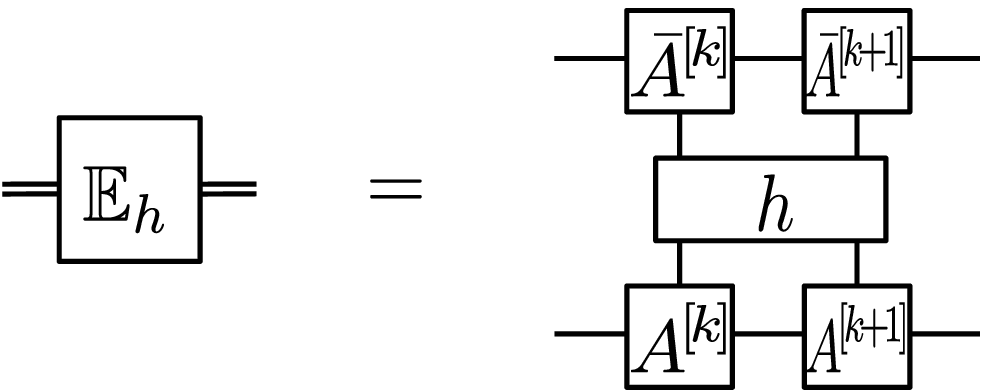}}
\end{equation}
and 
\[
\raisebox{-0.9cm}{
\includegraphics[scale=0.55]{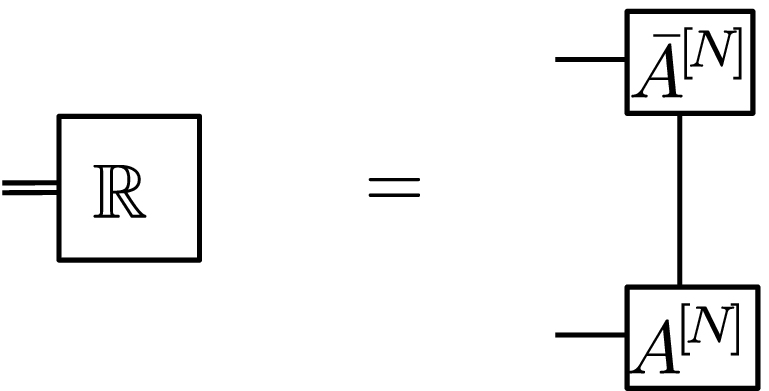}}\quad .
\]
All of these operators can be computed efficiently (in the parameters $D$
and $N$), as they are vectors/matrices of fixed dimension $D^2$, and can
be obtained by contracting a constant number of indices.

Using the newly defined objects $\mathbb L$, $\mathbb E$, $\mathbb E_h$,
and $\mathbb R$, the expectation value $\bra\Psi h\ket\Psi$,
Eq.~(\ref{eq:exp-value-tn}), can be rewritten as
\begin{align*}
\bra\Psi h \ket\Psi & = 
\mathbb L\mathbb E^{[2]}\cdots \mathbb E^{[k-1]}
    \mathbb E_h \mathbb E^{[k+2]}\cdots \mathbb R
\\
&= \raisebox{-0.3cm}{
    \includegraphics[scale=0.55]{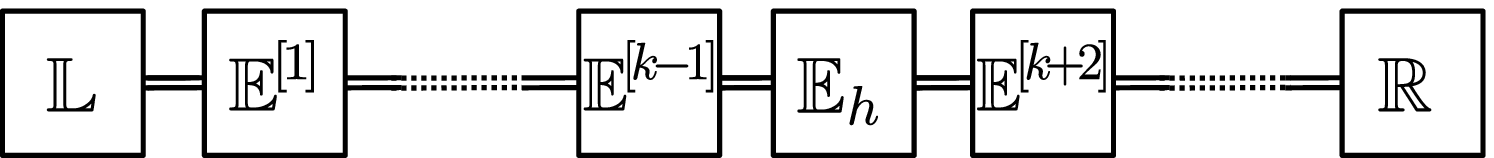}}\quad .
\end{align*}
This is, $\bra\Psi h\ket\Psi$ can be computed by multiplying
a $D^2$-dimensional vector $O(N)$ times with $D^2\times D^2$ matrices.
Each of these multiplication takes $O(D^4)$ operations, and thus,
$\bra\Psi h\ket\Psi$ can be evaluated in $O(ND^4)$ operations. There are
$O(N)$ terms in the Hamiltonian, and thus, the energy $\bra\Psi \sum_i
h_i \ket\Psi/\langle\Psi\ket\Psi$ can be evaluated in time $O(N^2D^4)$,
and thus efficiently;  in fact, this method can be easily improved to
scale as $O(ND^3)$.\footnote{Firstly, one uses that the products
$\mathbb L \cdots \mathbb E^{[s]}$ and $\mathbb E^{[s]}\cdots \mathbb R$
need to be computed only once (this can be simplified even further by
choosing the appropriate
gauge~\cite{schollwoeck:rmp,perez-garcia:mps-reps}), reducing the
scaling in $N$ to $O(N)$.  Secondly, one slightly changes the contraction
order: Starting from the
left, one contracts $A^{[1]}$, $\bar A^{[1]}$, $A^{[2]}$, $\bar A^{[2]}$,
$A^{[3]}$, etc.: This involves multiplications of $D\times D$ matrices
with $D\times dD$ matrices, and $D\times Dd$ matrices with $Dd\times D$
matrices, yielding a $O(dD^3)$ scaling.}  Similarly, one can see that
e.g.\ correlation functions $\bra\Psi P_i\otimes Q_j\ket\Psi$ or string
order parameters $\bra\Psi X\otimes X\otimes \cdots \otimes X\ket\Psi$ can
be reduced to matrix multiplications and thus evaluated in $O(ND^3)$. 
Exactly the same way, evaluating expectation values for MPS with
periodic boundary conditions can be reduced to computing the trace of a
product of matrices $\mathbb E$ of size $D^2\times D^2$. Each
multiplication scales like $O(D^6)$, and using the same tricks as before,
one can show that for systems with periodic boundary conditions,
expectation values can be evaluated in time $O(ND^5)$.

In summary, we find that energies, correlations functions, etc.\ can be
efficiently evaluated for MPS, with computation times scaling as $O(ND^3)$
and $O(ND^5)$ for open and periodic boundary conditions, respectively.

\subsubsection{Variational optimization of MPS}

As we have seen, we can efficiently compute the energy of an MPS with
respect to a given local Hamiltonian $H=\sum_i h_i$.  In order to use MPS
for numerical simulations, we still need to figure out an efficient way to
find the MPS which minimizes the energy for a given $D$. To this end, let
us first pick a site $k$, and try to minimize the energy as a function
of $A^{[k]}$, while keeping all other MPS tensors $A^{[s]}$, $s\ne k$,
fixed.  Now, since $\ket\Psi$ is a linear function of $A^{[k]}$,
we have that
\[
\frac{\bra\Psi H\ket\Psi}{\langle\Psi\ket\Psi}
= \frac{\vec A^{[k]\dagger} X \vec A^{[k]}}{
	\vec A^{[k]\dagger} Y \vec A^{[k]}}
\]
is the ratio of two quadratic forms in $A^{[k]}$.
Here, $\vec A^{[k]}$ denotes the vectorized version of
$A^{[k]}_{i,\alpha\beta}$, where $(i,\alpha,\beta)$ is interpreted as a
single index. The matrices $X$ and $Y$ can be obtained by contracting the
full tensor network (\ref{eq:exp-value-tn}) except for the tensors
$A^{[k]}$ and $\bar A^{[k]}$, which can be done efficiently.
The $\vec A^{[k]}$ which minimizes this energy can be found
by solving the generalized eigenvalue equation
\[
X\vec A^{[k]} = E\,Y\vec A^{[k]}
\]
where $E$ is the energy;
again, this can be done efficiently in $D$. For MPS with open boundary
conditions, we can choose a gauge\footnote{
MPS have a natural gauge degree of freedom, since for any 
$X_s$ with a right inverse $X_s^{-1}$, we can always replace
\begin{align*}
A^{[s]}_i & \leftrightarrow \ A^{[s]}_i X_s\\
A^{[s+1]}_i & \leftrightarrow \ X_s^{-1} A^{[s+1]}_i 
\end{align*}
without changing the state; this gauge degree of freedom can be used to
obtain standard forms for MPS with particularly nice
properties~\cite{fannes:FCS,perez-garcia:mps-reps}.
}
for the tensors such that
$Y=\openone$~\cite{schollwoeck:rmp,perez-garcia:mps-reps} -- this
reduces the problem to a usual eigenvalue problem, and avoids problems due
to ill-conditioned $Y$.

This shows that we can efficiently minimize the energy as a function of
the tensor $A^{[k]}$ at an individual site $k$.  In order to minimize the
overall energy, we start from a randomly chosen MPS, and then sweep 
through the sites, sequentially optimizing the tensor at each site.
Iterating this a few times over the system (usually sweeping back and forth)
quickly converges to a state with low energy.  Although in principle, such
an optimization can get stuck~\cite{eisert:DMRG-NP,schuch:mps-gap-np}, in
practice it works extremely well and generally converges to the optimal
MPS (though some care might have to be put into choosing the initial
conditions).

In summary, we find that we can use MPS to efficiently simulate ground
state properties of one-dimensional quantum systems with both open and
periodic boundary conditions.  This simulation method can be understood as
a reformulation of the Density Matrix Renormalization Group (DMRG)
\index{Density Matrix Renormalization Group (DMRG)}
algorithm~\cite{white:DMRG,verstraete:dmrg-mps}, which is a renormalization
algorithms based  on keeping the states which are most relevant for the
entanglement of the system, and which since its invention has been highly
successful in simulating the physics of one-dimensional quantum systems
(see
Refs.~\cite{schollwoeck:rmp,schollwoeck:review-annphys,murg:peps-review}
for a review of the DMRG algorithm and its relation to MPS, and a more
detailed discussion of MPS algorithms).

\subsection{MPS as a framework for solvable models}

Up to now, we have seen that MPS form a powerful framework for the
simulation of one-dimensional systems, due to their ability to efficiently
approximate ground states of local Hamiltonians, and due to the fact that
the necessary variational minimization can be carried out efficiently.  As
we will see in the following, MPS can also be used as a framework to
construct solvable models whose properties can be understood analytically,
which makes them a powerful tool for resolving various analytical
problems regarding quantum many-body systems.

\subsubsection{The AKLT model and parent Hamiltonians
\label{sssec:aklt-and-parent}}

\begin{figure}
\centering
\includegraphics[height=3.3cm]{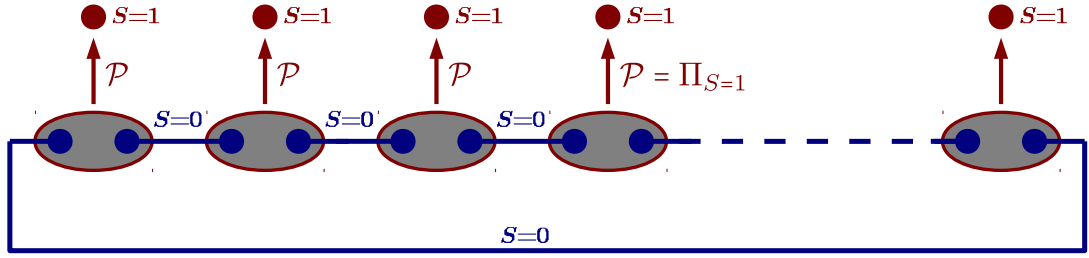}
\caption{\label{fig:aklt}
The AKLT model is built from spin-$\tfrac12$ singlets ($S=0$) between
adjacent sites, which are projected onto the spin one subspace ($\mc
P=\Pi_{S=1}$) as indicated.}
\end{figure}

The paradigmatic analytical MPS model is the so-called AKLT state, named
after Affleck, Kennedy, Lieb, and Tasaki~\cite{affleck:aklt-prl,aklt}. The construction of
the translational invariant AKLT state is
illustrated in Fig.~\ref{fig:aklt}: The auxiliary systems are spin-$\tfrac12$
particles, which are put into a singlet (i.e., spin $S=0$)
state\footnote{This bond can be easily replaced by a normal bond
$\ket{00}+\ket{11}$ by applying the transformation
$\left(\begin{smallmatrix}0&1\\-1&0\end{smallmatrix}\right)$ on one side,
which can subsequently be absorbed in the map $\mc P$.}
$\ket{\omega}=\ket{01}-\ket{10}$. The two virtual spin-$\tfrac12$'s at
each site have jointly spin $\tfrac12\otimes \tfrac12=0\oplus 1$, and the
map $\mc P$ projects onto the spin-$1$ subspace, thereby describing a
translationally invariant chain of spin-$1$ particles.

\begin{figure}[t]
\centering
\includegraphics[height=2.5cm]{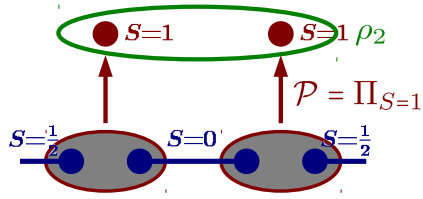}
\caption{\label{fig:aklt-rho2}
Construction of the parent Hamiltonian for the AKLT model.}
\end{figure}

Let us now see what we can say about the reduced density operator $\rho_2$
of two consecutive sites  in the AKLT model. The following argument is
illustrated in Fig.~\ref{fig:aklt-rho2}: On the virtual level, we start from two
``open'' bonds, each of which has spin $\tfrac12$ (while we might have
more information about their joint state, we are free to neglect it), as
well as one ``interior'' bond which has spin $0$; thus, the virtual state
we start from has spin 
\[
\tfrac12\otimes0\otimes\tfrac12 = 0\oplus 1\ .
\]
The projections $\mc P$ onto the spin $1$ subspace, on the other hand, do
not change the total spin (they only affect the weight of different
subspaces), which implies that $\rho_2$ has spin $0$ or $1$ as well. On
the other hand, $\rho_2$ is a state of two physical sites, each of which
has spin $1$; this is, it could have spin $1\otimes 1=0\oplus1\oplus2$.
We therefore find that there is a non-trivial constraint on $\rho_2$
arising from the AKLT construction: $\rho_2$ cannot have spin $2$.

We can now use this to construct a non-trivial Hamiltonian which has the
AKLT state as its ground state. Namely, let $h_{i,i+1}=\Pi_{S=2}$ be a
local Hamiltonian acting on sites $i$ and $i+1$, projecting onto the spin
$S=2$ subspace of the two sites.  Then, by the preceding argument, 
$h_{i,i+1}\ket{\Psi_\mathrm{AKLT}} = 0$
with $\ket{\Psi_\mathrm{AKLT}}$ the AKLT state, and thus, with
\[
H=\sum_i h_{i,i+1}\ ,
\]
we have that
\[
H\ket{\Psi_\mathrm{AKLT}} = 0\ .
\]
On the other hand, $H\ge0$, and thus, the AKLT state is a ground state of
the AKLT Hamiltonian $H$.

While the argument we have given right now seems specific to the AKLT
model, it turns out that the very same way, we can construct a so-called
``parent Hamiltonian'' for any MPS.  To this end, note that the reduced
density operator $\rho_k$ for $k$ consecutive sites lives in a $d^k$
dimensional space, yet it can have rank at most $D^2$ (since the only free
parameters we have are the outermost ``open'' bonds). Thus, as soon as
$d^k>D^2$, $\rho_k$ does not have full rank, and we can build a local
Hamiltonian $h$ acting on $k$ consecutive sites as
$h=\openone-\Pi_{\mathrm{supp}(\rho_k)}$ which has the given MPS as its
ground state.

Of course, it is not sufficient to have a non-trivial Hamiltonian with the
AKLT state (or any other MPS) as its ground state, but we want that the
AKLT state is the unique ground state of this Hamiltonian.  Indeed, it
turns out that there is a simple condition which is satisfied for almost
any MPS (and in particular for the AKLT state) which allows to prove that
it is a unique ground state of a suitable chosen parent
Hamiltonian; for the case of the AKLT model, it turns out that
this Hamiltonian is the two-body Hamitonian which we defined 
originally.\footnote{This condition is known as ``injectivity'' and
states there is a $k$ such that after blocking $k$ sites, the map
$X\mapsto \sum\mathrm{tr}[A_{i_1}\cdots A_{i_k}X]\ket{i_1,\dots,i_k}$ from
the ``open bonds'' to the physical system is injective, or, differently
speaking, that we can infer the state of the bonds by looking at the
physical system; the AKLT state acquires injectivity for $k=2$. It can
then be shown that the parent Hamiltonian defined on $k+1$ sites has a
unique ground state; in the case of the AKLT model, one can additionally
show that the ground space of this three-body parent Hamiltonian and the
original two-body one are
identical~\cite{aklt,fannes:FCS,perez-garcia:mps-reps,schuch:peps-sym,schuch:rvb-kagome}.}
Secondly, it can be shown that the parent Hamiltonian of MPS is always
gapped, this is, there is a finite gap above the ground space which does
not close even as $N\rightarrow\infty$.

Together, this proves that the AKLT state is the unique ground state of a
gapped local Hamiltonian.  It is an easy exercise to show that the AKLT
Hamiltonian equals 
\[
h_{i,i+1}=\tfrac12\bm S_i\cdot \bm S_{i+1} + \tfrac16(\bm S_i\cdot \bm
S_{i+1})^2 + \tfrac13\ ,
\]
and is thus close to the spin-$1$ Heisenberg model.  Indeed, proving that
the spin-$1$ Heisenberg model is gapped, as conjectured by Haldane, was
one motivation for the construction of the AKLT model, showing that MPS
allow for the construction of models of interest for which certain
properties can be rigorously proven.

The fact that MPS naturally appear as ground states of local Hamiltonians
makes them a very powerful tool in studying the properties of
one-dimensional gapped systems. In particular, one can use them to
classify phases of one-dimensional systems: On the one hand, we know that
the ground state of any gapped one-dimensional Hamiltonian is well
approximated by an MPS.  This allows us to replace the original
Hamiltonian by the parent Hamiltonian of the MPS while keeping a gap
(i.e., without crossing a phase transition).  Given two MPS characterized
by tensors $\mc P_1$ and $\mc P_2$, we can now build a smooth
interpolation between the two models (i.e., Hamiltonians) by appropriately
interpolating between $\mc P_1$ and $\mc P_2$:  due to the very way the
parent Hamiltonian is constructed, this will give rise to a smooth path of
parent Hamiltonians as well~\cite{schuch:mps-phases}. This is particularly
interesting if one imposes symmetries on the Hamiltonian, which (for a
system with a unique MPS ground state $\ket\Psi$) implies that $U_g^{\otimes
N}\ket\Psi = \ket\Psi$.  For an MPS with such a symmetry, it can be
shown~\cite{perez-garcia:stringorder-1d,sanz:mps-syms} that the symmetry
is reflected in a \emph{local} symmetry of the MPS map $\mc P$,
$U_g \mc P = \mc P (V_g\otimes \bar V_g)$, and classifying the possible
inequivalent types of $V_g$ allows to classify the different phases in
one-dimensional systems in the presence of
symmetries~\cite{pollmann:1d-sym-protection-prb,chen:1d-phases-rg,schuch:mps-phases}.

\subsubsection{Correlation length and the transfer operator}

As we have discussed in the preceding section, MPS form a powerful tool
for the construction of solvable models.  Let us now have a closer look
at what we can say about the behavior of physical quantities. While
we have discussed how to extract physically relevant quantities such as
energies and correlation functions numerically in
Sec.~\ref{ssec:compute-expval}, the purpose of this section is different:
Here, we focus on a translationally invariant exact MPS
ansatz, and we are asking for simple (and ideally closed) expressions for
quantities such as the correlation length.

In Sec.~\ref{ssec:compute-expval}, the central object was the
transfer operator,
\[
\raisebox{-0.9cm}{
\includegraphics[scale=0.55]{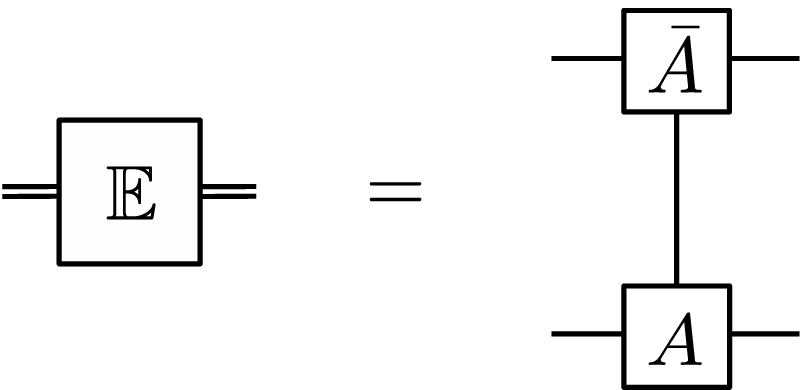}}\quad.
\]
As compared to Eq.~(\ref{eq:top-def-sitedep}), we have omitted the
site-dependency index $[s]$.  It is easy to see that -- up to local
unitaries -- the MPS is completely determined by the transfer operator
$\mathbb E$: If $A$ and $B$ give rise to the same transfer operator,
$\sum A_i\otimes \bar A_i = \sum B_i\otimes \bar B_i$, then we must have
$A_i = \sum_j v_{ij} B_j$, with $V=(v_{ij})$ an isometry.\footnote{This
is nothing but the statement that two purifications of a mixed state are
identical up to an isometry on the purifying system (this can be easily
proven using the Schmidt decomposition, cf.~Sec.~2.5 of
Ref.~\cite{nielsen-chuang}), where the index $i$ corresponds to the
purifying system.} In the following, we would like to see how specific
properties can be extracted from the transfer operator.

One property which is of particular interest are two-point correlations.
To obtain those, let us first define 
\[
\raisebox{-0.9cm}{
\includegraphics[scale=0.55]{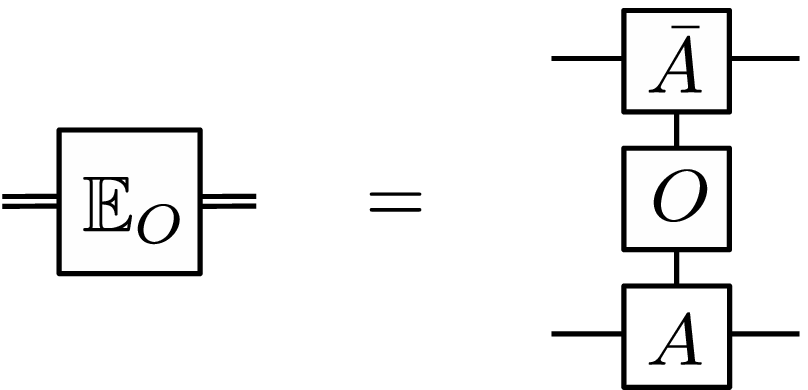}}\quad 
=\ \sum_{ij} \bra{j}O\ket{i}A_i\otimes\bar A_j\ .
\]
Then, the correlation between operator $A$ at site $1$ and $B$ at site
$j+2$ on a chain of length $N\rightarrow\infty$
is
\begin{equation}
\label{eq:corrfun-1}
\langle A_1B_{j+2}\rangle = 
\frac{\mathrm{tr}[\mathbb E_A \mathbb E^j \mathbb E_B\mathbb E^{N-j-2}]}{
\mathrm{tr}[\mathbb E^N]}\ .
\end{equation}
If we now assume that $\mathbb E$ has a unique maximal eigenvalue
$\lambda_1$, which
we normalize to $\lambda_1=1$, with eigenvectors $\ket{r}$ and $\bra{l}$, then
for large $N$, $\mathbb E^N\rightarrow \ket r\bra l$, and thus
Eq.~(\ref{eq:corrfun-1}) converges to $\bra l \mathbb E_A \mathbb E^j
\mathbb E_B\ket r$.  If we now expand
\[
E^j = \ket r\bra l + \sum_{k\ge 2} \lambda_k^j \ket{r_k}\bra{l_k}
\]
(where $|\lambda_k|<1$ for $k\ge2$), we find that 
\[
\langle A_1B_{j+2}\rangle = 
\bra l \mathbb E_A \ket r\bra l \mathbb E_B\ket l + 
    \sum_{k\ge 2} \lambda_k^j 
    \bra l \mathbb E_A \ket{r_k}\bra{l_k}\mathbb E_B\ket l\ ,
\]
and thus decays exponentially with $j$; in particular, the correlation
length $\xi$ is given by the ratio of the second largest to the largest
eigenvalue of $\mathbb E$,
\[
\xi = -\frac{1}{\mathrm{ln}\big|\tfrac{\lambda_2}{\lambda_1}\big|}\ ,
\]
i.e., it is determined solely by simple spectral properties of the
transfer operator.  Let us note that the same quantity, i.e.\ the ratio
between the two largest eigenvalues, also bounds the gap of the parent
Hamiltonian~\cite{fannes:FCS,nachtergaele:degen-mps}.

The proof above that correlations in MPS decay exponentially is not
restricted to the case of a unique maximal eigenvalue of the transfer
operator: In case it is degenerate, this implies that the correlation
function has a long-ranged part (i.e., one which is not decaying), such as
in the GHZ state, while any other contribution still decays exponentially.
In particular, MPS cannot exhibit algebraically decaying correlations,
and in case MPS are used to simulate critical systems, one has
to take into account that a critical decay of correlations in MPS
simulations is always achieved by approximating it by a sum of
exponentials.

\section{Tensor network states beyond one dimension and ground state
problems
\label{sec:2d-and-beyond}
}

\subsection{PEPS: Tensor networks in two dimensions}

As we have seen, MPS are very well suited for simulating ground state
properties of one-dimensional systems.  But what if we want to go beyond
one-dimensional systems, and, e.g., study interacting spin systems in two
dimensions? Two-dimensional systems can exhibit a rich variety of
phenomena, such as  topologically ordered
states~\cite{wen:book,kitaev:toriccode} which are states distinct from
those in the trivial phase, yet which do not break any (local) symmetry.
Moreover,
two-dimensional spin systems can be highly frustrated due to the presence
of large loops in the
interaction graph, and even classical two-dimensional spin glasses can be
hard to solve~\cite{barahona:spinglass-np}. In the following, we will
focus on the square lattice without loss of generality.

\begin{figure}
\centering
\includegraphics[scale=0.55]{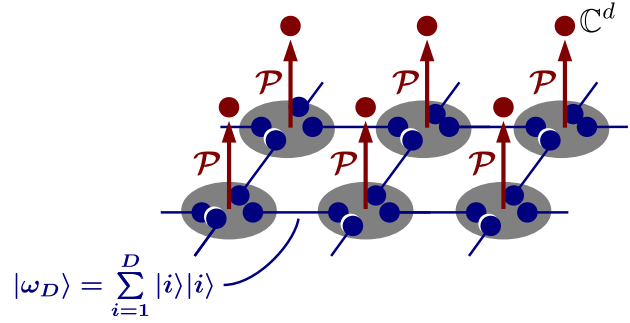}
\caption{\label{fig:peps-constr}
PEPS construction for a 2D square lattice, where we have omitted the
site-dependence $\mc P\equiv \mc P_s$ of the maps $\mc P_s$.}
\end{figure}

A first idea to simulate two-dimensional systems would be to simply use an
MPS, by choosing a one-dimensional ordering of the spins in the
two-dimensional lattice.  While this approach has been applied
successfully (see, e.g., Ref.~\cite{yan:heisenberg-kagome}), it cannot
reproduce the entanglement features of typical ground states in two
dimensions as
one increases the system size: As we have discussed in
Section~\ref{ssec:arealaw},  two-dimensional systems also satisfy an area
law, i.e., in the ground state we expect the entanglement of a region $A$
with its complement to scale like its boundary, $S(\rho_A)\sim |\partial
A|$.  To obtain an ansatz with such an entanglement scaling, we follow the
same route as in the construction of MPS: We consider each site as being 
composed of \emph{four} $D$-dimensional virtual subsystems, place each of
them in a maximally entangled state $\ket{\omega_D}$ with the
corresponding subsystem of each of the adjacent sites, and finally apply a
linear map 
\[
\mc P_s:\mathbb C^D\otimes \mathbb C^D\otimes \mathbb C^D\otimes
\mathbb C^D \rightarrow \mathbb C^d
\]
at each site $s$ to obtain a description of the physical state on a 2D
lattice of $d$-level sytems. The construction is illustrated in
Fig.~\ref{fig:peps-constr}.
Due to the way they are constructed, these states are called
\emph{Projected Entangled Pair States} (PEPS).\index{Projected Entangled
Pair States (PEPS)} Again, we can define 
five-index tensors $A^{[s]}=A^{[s]}_{i,\alpha\beta\gamma\delta}$, where
now 
\[
\mc P_s=\sum_{i\alpha\beta\gamma\delta}
    A^{[s]}_{i,\alpha\beta\gamma\delta}
    \ket{i}\bra{\alpha,\beta,\gamma,\delta}\ ,
\]
and express the PEPS in Fig.~\ref{fig:peps-constr} graphically as a tensor network
\[
\raisebox{-0.6cm}{\includegraphics[scale=0.8]{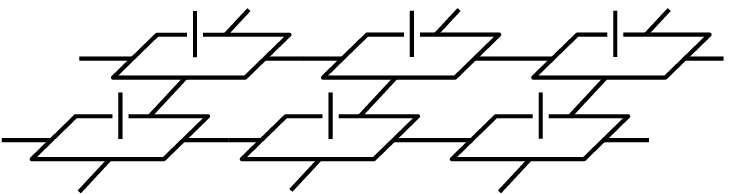}}
\]
(where we have omitted the tensor labels). Similar to the result in
one dimension, one can show that PEPS approximate ground states of local
Hamiltonians well as long as the density of states grows at most
polynomially with the
energy~\cite{hastings:locally,hastings:mps-entropy-ent}, and thereby
provide a good variational ansatz for two-dimensional systems. (Note,
however, that it is not known whether all 2D states which obey an area law
are approximated well by PEPS.)

\subsubsection{\label{ssec:peps-contr}
Contraction of PEPS}

Let us next consider what happens if we try to compute expectation values
of local observables for PEPS. For simplicity, we first discuss the
evaluation of the normalization $\langle \Psi\ket\Psi$, which is obtained by
sandwiching the ket and bra tensor network of $\ket\Psi$,
\begin{equation}\label{eq:peps-normalization}
\langle\Psi\ket\Psi\quad=\qquad\raisebox{-1.5cm}{\includegraphics[scale=0.55]{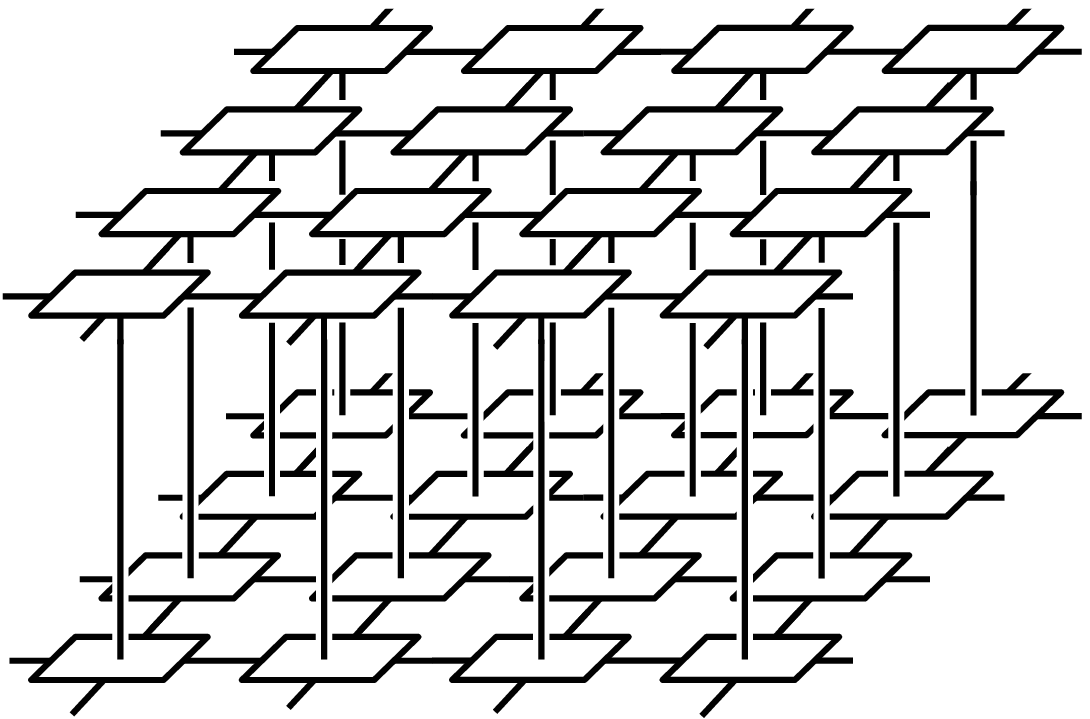}}\qquad .
\end{equation}
This can again be expressed using transfer operators
\index{transfer operator}
\[
\includegraphics[scale=0.55]{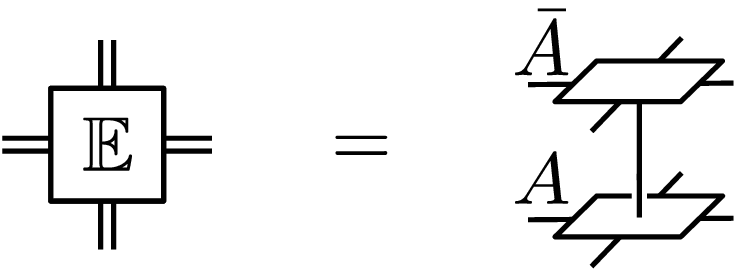}
\]
(the $\mathbb E$ should be thought of as being ``viewed from the top''),
leaving us with the task of contracting the network
\[
\raisebox{-1.5cm}{\includegraphics[scale=0.55]{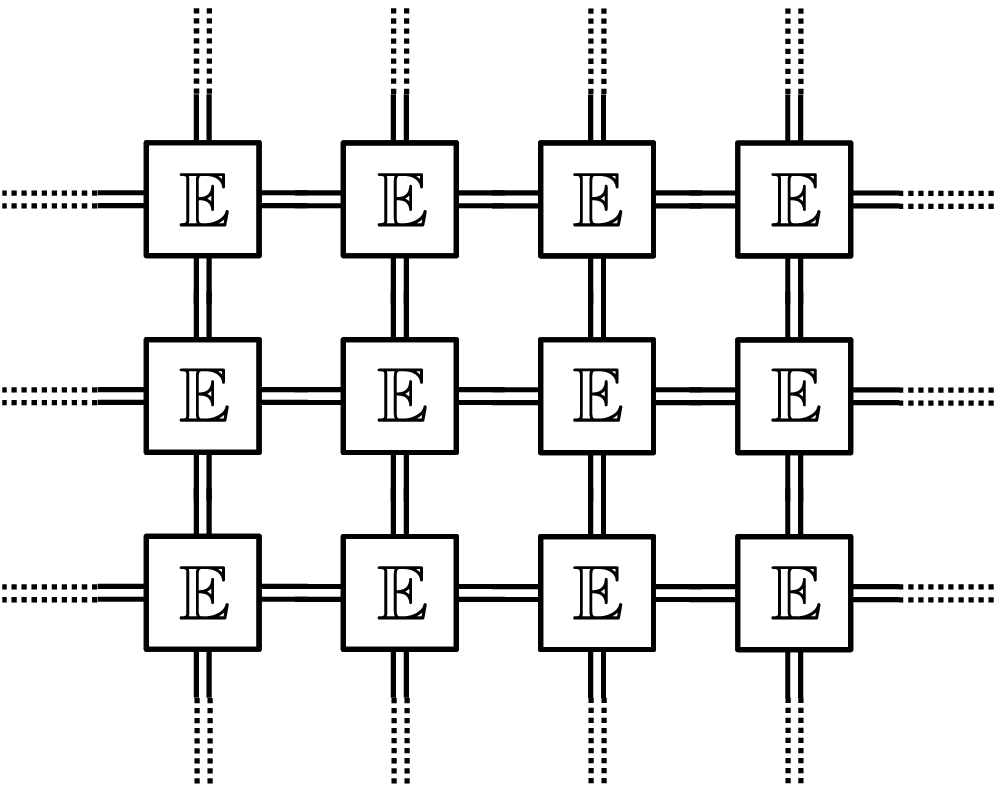}}\qquad .
\]
[This easily generalizes to the computation of expectation values, where
some of the $\mathbb E$ have to be modified similar to
Eq.~(\ref{eq:def-top-with-h})].  Different from the case of MPS, there is
no one-dimensional structure which we can use to reduce this problem to
matrix multiplication.  In fact, it is easy to see that independent of the
contraction order we choose, the cluster of tensors we get (such as a
rectangle) will at some point have a boundary of a length comparable to
the linear system size. This is, we need to store an object with a number
of indices proportional to $\sqrt N$ -- and thus an \emph{exponential}
number of parameters -- at some point during the contraction, making it
impossible to contract such a network efficiently.  (Indeed, it can be
proven that such a contraction is a computationally hard
problem~\cite{schuch:cplx-of-PEPS}.)

This means that if we want to use PEPS for variational calculations in two
dimensions, we have to make use of some approximate contraction scheme,
which of course should have a small and ideally controlled error.
\index{approximate contraction} To this end, we proceed as
follows~\cite{verstraete:2D-dmrg}: Consider the contraction of a
two-dimensional PEPS with open boundary conditions,
\begin{equation}\label{eq:approx-contr-startTN}
\raisebox{-2.2cm}{\includegraphics[scale=0.55]{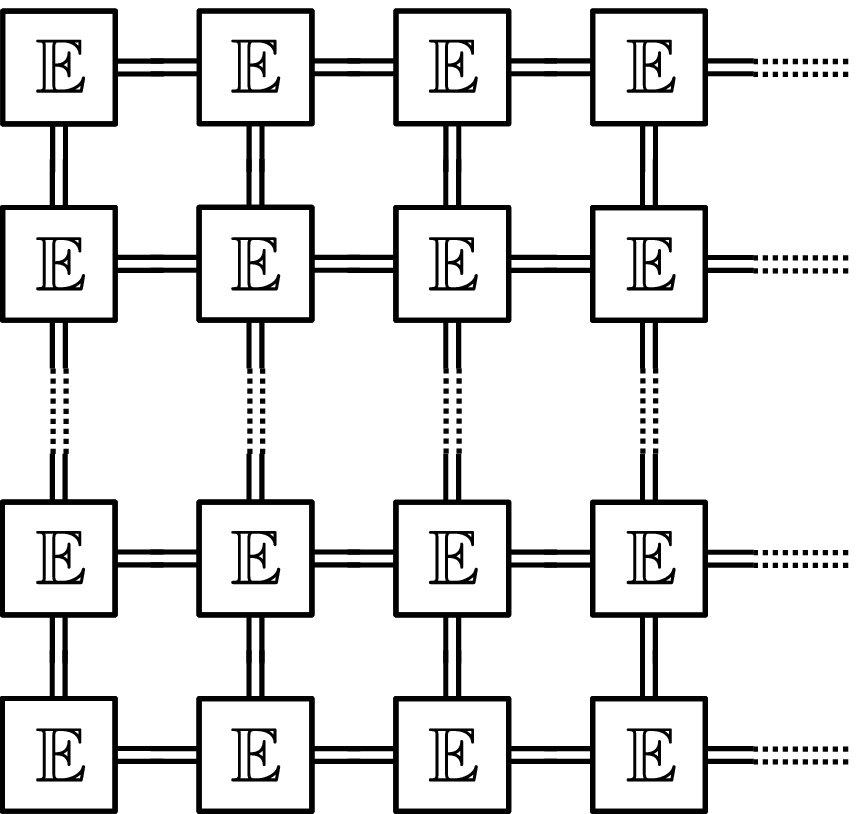}}\qquad .
\end{equation}
Now consider the first two columns, and block the two tensors in each
column into a new tensor $\mathbb F$ (with vertical bond dimension $D^4$):
\begin{equation}\label{eq:approx-contr-twocol}
\raisebox{-2.2cm}{\includegraphics[scale=0.55]{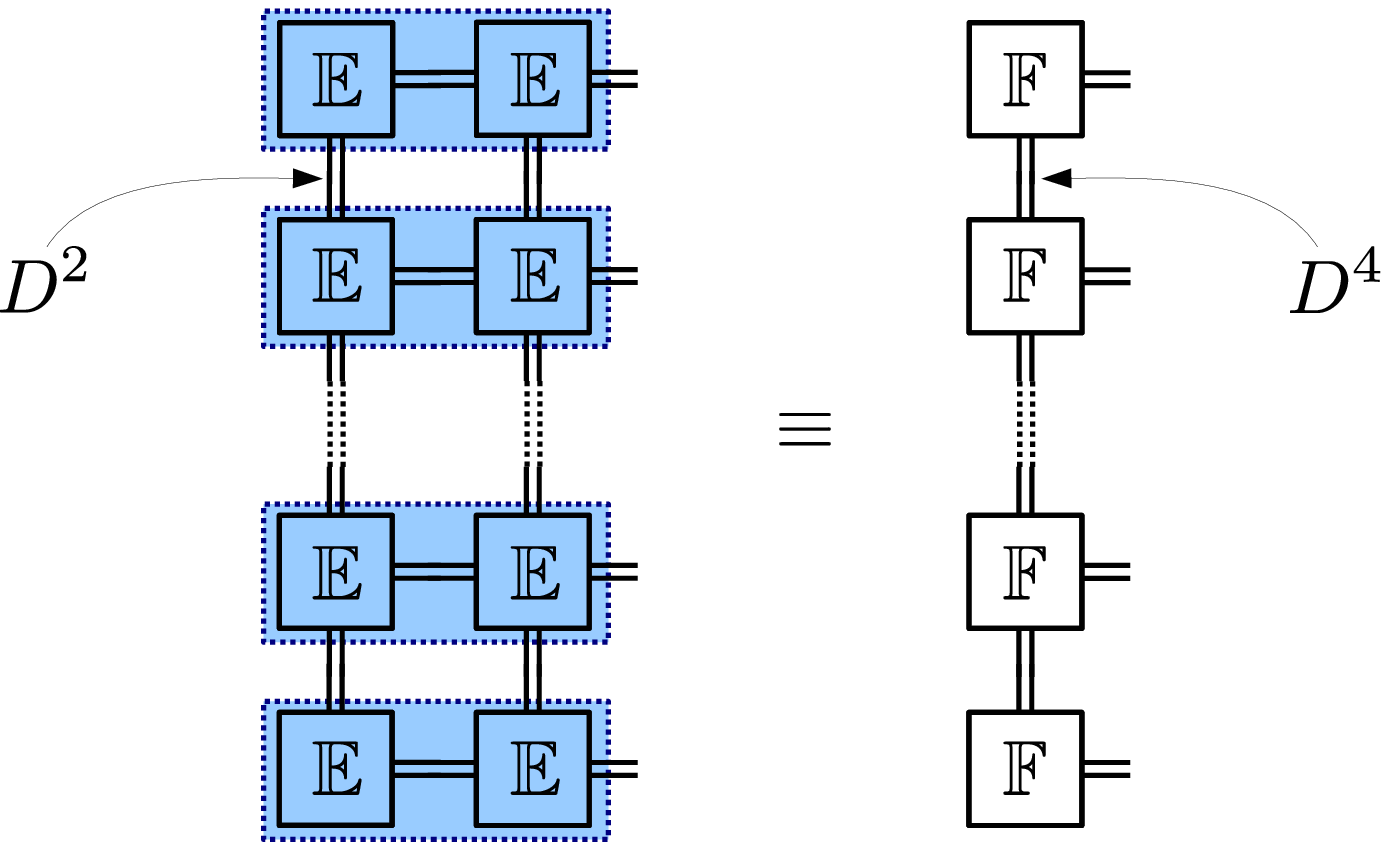}}\qquad .
\end{equation}
This way, we have reduced the number of columns in
(\ref{eq:approx-contr-startTN}) by one.  Of course, this came at the cost
of squaring the bond dimension  of the first
column, so this doesn't help us yet.  However, what we do now is to
approximate the right hand side of (\ref{eq:approx-contr-twocol}) by an
MPS with a (fixed) bond dimension $\alpha D^2$ for some $\alpha$. We can
then iterate this procedure column by column, thereby contracting the
whole MPS, and at any point, the size of our tensors stays bounded.
It remains to be shown that the elementary step of approximating an MPS
$\ket\Phi$ [such as the r.h.s.\ of (\ref{eq:approx-contr-twocol})] by an
MPS $\ket\Psi$ with smaller bond dimension can be done efficiently: To
this end, it is sufficient to note that the overlap $\langle\Phi\ket\Psi$
is linear in each tensor $A^{[s]}$ of $\ket\Psi$, and thus, maximizing the
overlap 
\[
\frac{\big|\langle\Phi\ket\Psi\big|^2}{\langle\Psi\ket\Psi}
\]
can again be reduced to solving a generalized eigenvalue problem, just as
the energy minimization for MPS in the one-dimensional variational method.
Differently speaking, the approximate contraction scheme succeeds by
reducing the two-dimensional contraction problem to a sequence of
one-dimensional contractions, i.e., it is based on a dimensional reduction
of the problem.

This shows that PEPS can be contracted approximately in an efficient
way. 
The scaling in $D$ is naturally much less favorable
than in one dimension, and for the most simple approach one finds a
scaling of $D^{12}$ for open boundaries, which using several tricks can be
improved down to $D^8$.  Yet, the method is limited to much smaller $D$ as
compared to the MPS ansatz. It should be noted that the approximate
contraction method we just described has a controlled error,
as we know the error made in in each
approximation step.  Indeed, the approximation is very accurate as long
as the system is short-range correlated, and the accuracy of the method is
rather limited by the $D$ needed to obtain a good enough approximation of
the ground state.
  Just as in one dimension, we can
use this approximate contraction method to build a variational method for
two-dimensional systems by successively optimizing over individual
tensors~\cite{verstraete:2D-dmrg}.

\subsubsection{Alternative methods for PEPS calculations\label{ssec:peps-ext}}

The PEPS construction is not limited to square lattices, but can be
adapted to other lattices, higher dimensions, and even  arbitrary
interaction graphs.  Clearly, the approximate contraction scheme we just
presented works for any two-dimensional lattice, and in fact for any
planar graph.  In order to approximately contract systems in more than two
dimensions, note that the approximate contraction scheme is essentially a
scheme for reducing the dimension of the problem by one; thus, in order to
contract e.g.\ three-dimensional systems we can nest two layers of the
scheme.  In cases with a highly connected PEPS
graph (e.g., when considering systems with highly connected interaction
graphs such as orbitals in a molecule), one can of course still try to
find a sequential contraction scheme, though other contraction methods
might be more promising.

The contraction method described in Section~\ref{ssec:peps-contr} is not
the only contraction scheme for PEPS. One alternative method is based on
renormalization ideas~\cite{jiang:2nd-terg,gu:TERGidea,xie:2nd-rg}:
There, one takes blocks of e.g.\ $2\times 2$
tensors and tries to approximate them by a tensor with lower bond
dimension by appropriate truncation,
\[
\raisebox{-1cm}{\includegraphics[scale=.55]{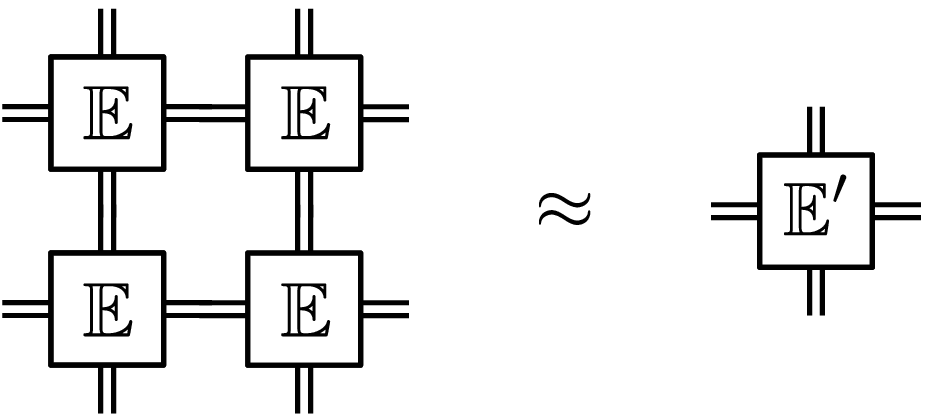}}\quad.
\]
Finding the best truncation scheme requires exact knowledge of the
environment, i.e., the contraction of the remaining tensor network.  Since
this is as hard as the original problem, heuristic methods to approximate
the environment (such as to only contract a small number of surronding
tensors exactly, and imposing some boundary condition beyond that) have
been introduced.  While these approximations are in principle less
accurate and the error is less controlled, their more favorable scaling
allows for larger $D$ and thus potentially better approximations of the
ground state. 

Another approach to speed up PEPS contraction is using Monte Carlo
sampling~\cite{schuch:sbs,sandvik:mps-mc,wang:mc-peps}: 
\index{Monte Carlo}
We can always
write
\begin{equation}\label{eq:mc-sampling}
\frac{\bra{\Psi}O\ket{\Psi}}{\langle\Psi\ket\Psi} = 
\sum_i
p(i) \frac{\bra i O \ket\Psi}{\langle i\ket\Psi}\ ,
\end{equation}
where the sum runs over an orthonormal basis $\ket{i}$, and $p(i)=|\langle
i\ket\Psi|^2/\langle\Psi\ket\Psi$; in particular, we want to consider the
local spin basis $i=(i_1,\dots,i_N)$.  If we can compute $\langle i\ket\Psi$ and
$\bra i O\ket\Psi$ (where the latter reduces to the former if $O$ is a local
operator), then we can use Monte Carlo sampling to approximate the
expectation value $\bra\Psi O\ket\Psi$. In particular, for PEPS $\langle
i\ket\Psi$ can again be evaluated by contracting a two-dimension tensor
network; however, this network now has bond dimension $D$ rather than
$D^2$.  Thus, we can apply any of the approximate contraction
schemes described before, but we can go to much larger $D$ with the same
computational resources; it should be noted, however, that the number 
of operations needs
to be multiplied with the number $M$ of sample points taken, and that the
accuracy of Monte Carlo sampling improves as $1/\sqrt{M}$.

\subsubsection{PEPS and exactly solvable models}

Just as MPS, PEPS are also very well suited for the construction of
solvable models.  First of all, many relevant states have exact PEPS
representations: For instance, the 2D cluster state underlying measurement
based computation~\cite{raussendorf:cluster-short} is a PEPS with
$D=2$~\cite{verstraete:mbc-peps}, as well as various topological models
such as Kitaev's toric code
state~\cite{kitaev:toriccode,verstraete:comp-power-of-peps,schuch:peps-sym} or Levin and
Wen's string-net
models~\cite{levin:stringnets,buerschaper:stringnet-peps,gu:stringnet-peps}.
A particularly interesting class of exact PEPS ansatzes is obtained by
considering classical spin models $H(s_1,\dots,s_N)$, such as the 2D Ising
model $H=-\sum_{<i,j>} s_is_j$, and defining a quantum state 
\[
\ket\psi = \sum_{s_1,\dots,s_N} e^{-\beta/2\,H(s_1,\dots,s_N)}
\ket{s_1,\dots,s_N}\ .
\]
It is easy to see that this state has the same correlation functions in
the $z$ basis as the classical Gibbs state at inverse temperature $\beta$.
On the other hand, this state has a PEPS representation (called the
``Ising PEPS'') with $\mc P =
\ket{0}\bra{a,a,a,a}+\ket{1}\bra{b,b,b,b}$, where $\langle a\ket a=\langle
b\ket b=1$ and $\langle a\ket b=\langle b\ket a =
e^{-\beta}$~\cite{verstraete:comp-power-of-peps}.  This implies that
in two dimensions, PEPS (e.g., the Ising PEPS at the critical $\beta$) can
exhibit algebraically decaying correlations which implies that any
corresponding Hamiltonian must be
gapless.\footnote{This follows from the so-called ``exponential clustering
theorem''~\cite{hastings:gap-and-expdecay,nachtergaele:exp-clustering}
which states that ground states of gapped Hamiltonians have exponentially
decaying correlation functions.}  This should be contrasted with the
one-dimensional case, where we have seen that any MPS has exponentially
decaying correlations and is the ground state of a gapped parent
Hamiltonians.

Clearly, parent Hamiltonians can be defined for PEPS just the same way as
for MPS: The reduced density operator $\rho_{k\times\ell}$ of a region of
size $k\times\ell$ lives on a $d^{k\ell}$-dimensional system, yet
can have rank at most $D^{2k+2\ell}$. Thus, by growing both $k$ and $\ell$
we eventually reach the point where $\rho_{k\times\ell}$ becomes rank
deficient, giving rise to a non-trivial parent Hamiltonian $H$ with local
terms $H$ projecting onto the kernel of $\rho_{k\times\ell}$.  Again, just as
in one dimension it is possible to devise conditions under which the
ground state of $H$ is unique~\cite{perez-garcia:parent-ham-2d}, as well
as modified conditions under which the ground space of $H$ has a
topological structure~\cite{schuch:peps-sym}, which turns out to be
related to ``hidden'' (i.e., virtual) symmetries of the PEPS tensors.
Unlike in one dimension, these parent Hamiltonians are not always gapped
(such as for the critical Ising PEPS discussed above); however, techniques
similar to the ones used in one dimension can be used to prove a gap for
PEPS parent Hamiltonians under certain
conditions~\cite{schuch:mps-phases}.

\subsection{Simulating time evolution and thermal states}

Up to now, our discussion has been focused on ground states of many-body
systems.  However, the techniques described here can also be adapted to
simulate thermal states as well as time evolution of systems governed by
local Hamiltonians.  In the following, we will discuss the implementation
for one-dimensional systems;  the generalization to to 2D and beyond is
straightforward.

Let us start by discussing how to simulate time evolution. 
\index{time evolution}
(This will
also form the basis for the simulation of thermal states.) We want to
study how an initial MPS $\ket\Psi$ changes under the evolution with
$e^{i H t}$; w.l.o.g., we consider $H$ to be nearest neighbor. To this
end, we perform a Trotter decomposition 
\index{Trotter decomposition}
\[
e^{i H t} \approx \left(e^{i H_\mathrm{even} t/M}e^{i H_\mathrm{odd}
t/M}\right)^M
\]
where we split $H=H_\mathrm{even}+H_\mathrm{odd}$ into even and odd terms
(acting between sites $12$, $34$, \dots, and $23$, $45$, \dots,
respectively), such that both $H_\mathrm{even}$ and $H_\mathrm{odd}$ are
sums of non-overlapping terms. For large $M$, the
Trotter expansion becomes exact, with the error scaling like $O(1/M)$.
We can now write
\[
e^{i H_\mathrm{even} \tau} = \bigotimes_{i=1,3,5,\dots} 
	e^{i h_{i,i+1}\tau} = 
\raisebox{-0.4cm}{\includegraphics[scale=0.6]{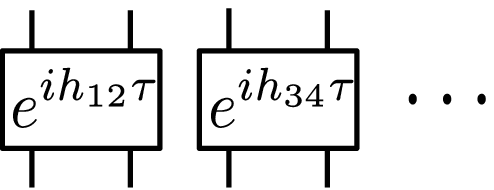}}
\]
(with $\tau = t/M$), and similarly for $e^{i H_\mathrm{odd}\tau}$. Thus,
after one time step $\tau$ the initial MPS is transformed into
\[
\raisebox{-0.2cm}{\includegraphics[scale=0.58]{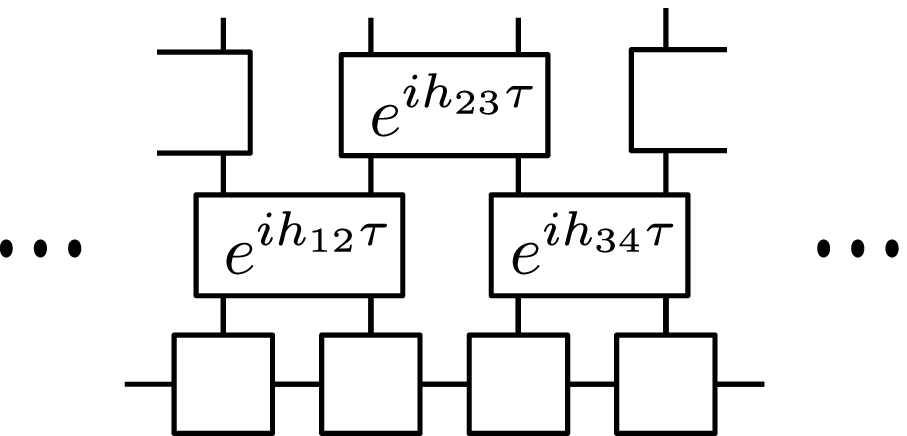}}\quad .
\]
Here, the lowest line is the initial MPS, and the next two lines the
evolution by $H_\mathrm{even}$ and $H_\mathrm{odd}$ for a time $\tau$,
respectively.  We can proceed this way and find that the state after a
time $t$ is described as
the boundary of  a two-dimensional tensor network. We can then use the
same procedure as for the approximate contraction of PEPS 
(proceeding row by row) to obtain an MPS 
description of the state at time $t$~\cite{vidal:timeevol}.  A caveat of
this method is that this only works well as long as the state has low
entanglement at all times, since only then, a good MPS approximation of
the state exists~\cite{verstraete:faithfully,schuch:mps-entropies}.  While
this holds for low-lying
excited states with a small number of quasiparticles, this is not true
after a quench, i.e., a sudden change of the overall Hamiltonian of the
system~\cite{cardy:ising,schuch:timeevol-hard}. However, this does not
necessarily rule out the possibility to simulate time evolution using
tensor networks, since in order to compute an expectation value $\bra\Psi
e^{-iHt}O e^{iHt}\ket\Psi$, one only needs to contract a two-dimensional
tensor network with no boundary, which can not only be done along the time
direction (row-wise) but also along the space direction (column-wise),
where such bounds on the correlations do not necessarily hold; indeed,
much longer simulations times have be obtained this
way~\cite{banuls:1d-timeevol-folding}.

In the same way as real time evolution, we can also implement imaginary
time evolution;\index{imaginary time evolution} and since $e^{-\beta H}$
acting on a random initial state approximates the ground state for
$\beta\rightarrow\infty$, this can be used as an alternative algorithm for
obtaining MPS approximations of ground states.

In order to simulate thermal states, 
\index{thermal states}
we use Matrix Product Density
Operators (MPDOs)~\cite{verstraete:finite-t-mps} \index{Matrix Product
Density Operator (MPDO)} 
\begin{align*}
\rho&=\sum_{ {i_1,\dots,i_N}\atop{j_1,\dots,j_N}}
    \tr[A^{[1]}_{i1,j1}\cdots A^{[N]}_{i_N,j_N}]
	\ket{i_1,\dots,i_N}\bra{j_1,\dots,j_N}\\
    &=\ \raisebox{-0.85cm}{\includegraphics[scale=.55]{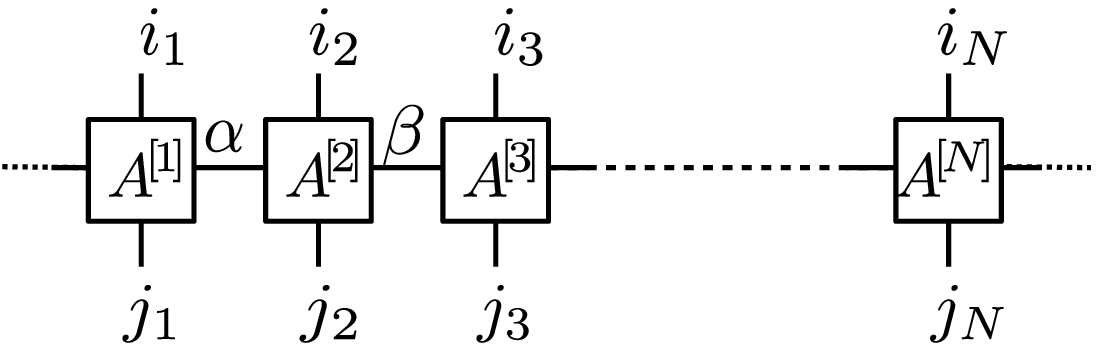}}\quad,
\end{align*}
where each tensor $A^{[s]}$ now has two physical indices, one for the ket
and one for the bra layer.  We can then write the thermal state as
\[
 e^{-\beta H} = e^{-\beta H/2} \openone e^{-\beta H/2}
\]
and use imaginary time evolution (starting from the maximally mixed state
$\openone$ which has a trivial tensor network description) and the Trotter
decomposition to obtain a
tensor network for $e^{-\beta H}$, which can again be transformed into an
MPDO with bounded bond dimension using approximate
contraction~\cite{verstraete:finite-t-mps}.

\subsection{Other tensor network ansatzes}

There is a number of other entanglement based ansatzes beyond MPS and PEPS
for interacting quantum systems, some of which we will briefly sketch in the
following.

\begin{figure}[t]
\centering
\includegraphics[scale=0.55]{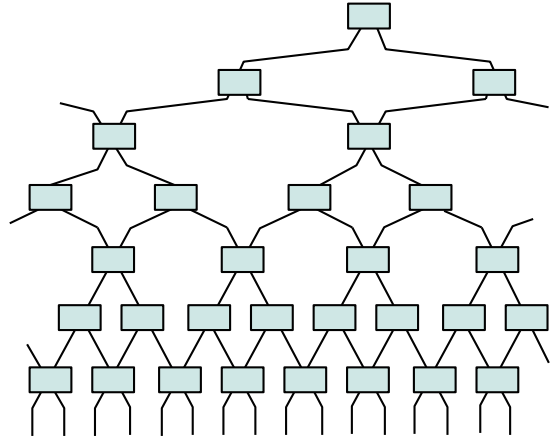}
\caption{\label{fig:mera}
The Multi-Scale Entanglement Renormalization Ansatz (MERA) in 1D. (The
left and right boundary are connected.)}
\end{figure}

Firstly, there is the Multiscale Entanglement Renormalization Ansatz
(MERA)~\cite{vidal:mera}, 
\index{Multiscale Entanglement Renormalization Ansatz (MERA)} 
which is an ansatz for scale invariant systems 
\index{scale invariance}
(these are systems at a
critical point where the Hamiltonian is gapless, and which have
algebraically
decaying correlation functions), and which incorporates the
scale-invariance in the ansatz. A first step towards a scale-invariant
ansatz would be to choose a tree-like tensor network.  However, such an
ansatz will not have sufficient entanglement between different blocks.
Thus, one adds additional \emph{disentanglers} which serve to remove the
entanglement between different blocks, which gives rise to the tensor network
shown in Fig.~\ref{fig:mera}. 
 In order to obtain an efficiently
contractible tensor network, one chooses the tensors to be
unitaries/isometries in vertical direction, such that each tensor cancels
with its adjoint. It is easy to see that this way for any local $O$, in the
tensor network for $\bra\Psi O\ket\Psi$ most tensors cancel, and one only
has to evaluate a tensor network of the size of the \emph{depth} of the
MERA, which is logarithmic in its length~\cite{vidal:mera}.  The MERA ansatz
is not restricted to one dimension and can also be used to simulate
critical system in 2D and beyond~\cite{evenbly:2d-mera}.

A different variational class is obtained by studying states for which
expectation values can be computed efficiently using Monte Carlo sampling.
\index{Monte Carlo} Following Eq.~(\ref{eq:mc-sampling}), this requires
(for local quantities $O$) that we can compute $\langle i\ket\Psi$
efficiently for all $i=(i_1,\dots,i_N)$. One class of states for which
this holds is formed by MPS, which implies that we can evaluate
expectation values for MPS using Monte Carlo
sampling~\cite{sandvik:mps-mc,schuch:sbs} (note that the scaling in $D$ is
more favorable since $\langle i\ket\Psi$ can be computed in time $\propto
ND^2$). This can be extended to the case where
$\langle i\ket\Psi$ is a product of efficiently computable objects, such
as products of MPS coefficients defined on subsets of spins: We can arrange
overlapping one-dimensional strings in a 2D geometry and associate to each
of them an MPS, yielding a class known as string-bond
states~\cite{schuch:sbs,sfondrini:sbs-3d},
\index{String-Bond States (SBS)}
 which combines a flexible geometry with the favorable scaling of MPS-based
methods.  We can also consider $\langle i\ket\Psi$ to be a product of 
coefficients each of which only depends on the spins $i_k$
supported on a small plaquette, and where the lattice is covered with
overlapping plaquettes, yielding a family of states known as Entangled
Plaquette States (EPS)~\cite{mezzacapo:EPS} 
\index{Entangled Plaquette States (EPS)}
or Correlator Product States (CPS)~\cite{changlani:cps},
\index{Correlator Product States (CPS)}
which again yields an efficient algorithm with flexible geometries.  In
all of these ansatzes, the energy is minimized by using a gradient method,
which is considerably facilitated by the fact that the gradient can be
sampled directly without the need to first compute the energy landscape.

In order to simulate infinite lattices, it is possible to extend MPS and
PEPS to work for infinite systems: iMPS and iPEPS.  
\index{infinite Matrix Product States (iMPS)}
\index{infinite Projected Entangled Pair States (iPEPS)}
The underlying idea is
to describe the system by an infinite MPS and PEPS with a periodic
pattern of tensors such as ABABAB\dots (which allows the system to break
translational symmetry and makes the optimization more well-behaved).
Then, one fixes all tensors except for one and minimizes the energy as a
function of that tensor until convergence is reached.  For the
optimization, one needs to determine the dependence of the energy on the
selected tensor, which can be accomplished in various ways, such as using
the fixed point of the transfer operator, renormalization methods
(cf.~Section~\ref{ssec:peps-ext}),  or the corner transfer matrix
approach.  For more information, see,
e.g.,~\cite{vidal:iTEBD,jordan:iPEPS,orus:iPEPS-CTM}.

\subsection{Simulation of fermionic sytems
\label{ssec:fermions}}

Up to now, we have considered the simulation of spin systems using tensor
networks.  On the other hand, in many cases of interest, such as for the
Hubbard model or the simulation of molecules, the underlying systems are
fermionic in nature.  In the following, we will discuss how tensor network
methods such as MPS, PEPS, or MERA can be extended to the simulation of
fermionic systems.  \index{fermionic tensor networks}

In order to obtain a natural description of fermionic systems, the idea is
to replace each object (i.e., tensor) in the construction of PEPS or MERA
by fermionic operators~\cite{corboz:fMERA,kraus:fPEPS,pineda:fMERA}.  This
is, in the construction of MPS and PEPS, Fig.~\ref{fig:mps-constr} and
Fig.~\ref{fig:peps-constr}, both the maximally entangled bonds and the
$\mc P_s$ are now built from fermionic operators and need to preserve
parity; equally, in the MERA construction, Fig.~\ref{fig:mera}, all
unitaries and isometries are fermionic in nature.  The resulting states are called
fermionic PEPS (fPEPS) \index{fermionic PEPS (fPEPS)} and \index{fermionic
MERA (fMERA)} fermionic MERA (fMERA).

Let us now have a closer look at a fermionic tensor network, and discuss
how to compute expectation values for those states.  E.g., the fPEPS
construction yields a state 
\[
(\mc P_1\otimes \mc P_2\otimes \cdots) (\omega_1\otimes \omega_2 \otimes
\cdots)\ket{\Omega}\ ,
\]
where $\ket{\Omega}$ is the vacuum state, the $\omega_i$ create entangled
fermionic states between the corresponding auxiliary modes, and the $\mc
P_s$ map the auxiliary fermionic modes to the physical fermionic modes at
site $s$ (leaving the auxiliary modes in the vacuum).  While the
product of the $\omega_i$ contains only auxiliary mode operators in a
given order, the product of the $\mc P_s$ contains the physical and
auxiliary operators for each site grouped together.  To compute
expectation values, on the other hand, we need to move all the physical
operators to the left and the virtual operators to the right in the
product of the $\mc P_s$; additionally, the virtual operators have to be
arranged such that they cancel with the ones arising from the product of
the $\omega_i$.  Due to the fermionic anti-commutation relations, this
reordering of fermionic operators results in an additional complication
which was not present for spin systems.  Fortunately, it
turns out that there are various ways how to take care of the ordering of
fermionic operators at no extra computational cost: One can use a
Jordan-Wigner transformation to transform the fermionic system to a spin
system~\cite{corboz:fMERA,pineda:fMERA}; one can map the fPEPS to a normal
PEPS with one additional bond which takes care of the fermionic
anticommutation relations~\cite{kraus:fPEPS}; or one can replace the
fermionic tensor network by a planar spin tensor network with parity
preserving tensors, where each crossing of lines [note that a planar 
embedding of a network such as the 2D expectation value in
Eq.~(\ref{eq:peps-normalization}) gives rise to crossings of lines, which
corresponds to the reordering of fermionic operators] is replaced by a
tensor which takes care of the anticommutation
rules~\cite{corboz:fMERA-crossingtensor,corboz:fPEPS-sim}.

\section{Quantum Complexity: Understanding the limitations to our
understanding
\label{sec:cplx}}

In the preceding sections, we have learned about different methods to
simulate the physics of many-body systems governed by local interactions,
using their specific entanglement structure. While these algorithms work
very well in practice, the convergence of these methods can almost never
be rigorously proven.  But does this mean that we just haven't found the
right proof yet, or is there a fundamental obstacle to proving that these
methods work -- and if yes, why do they work after all?

\subsection{A crash course in complexity classes}

These questions can be answered using the tools of complexity theory,
which deals with classifying the difficulty of problems. Difficulty is
typically classified by the resources needed 
to solve the problem on a computer
(such as computation time or
memory) as a function of the \emph{problem size}, this is, the length of
the problem description $n$.

Let us start by introducing some relevant complexity classes.\footnote{This is
going to be a very brief introduction. If you want to learn more, consult
e.g.~Refs.~\cite{papadimitriou:book,sipser:book}.} To simplify the
comparison of problems, complexity theorists usually focus on ``decision
problems'',
i.e., problems with a yes/no answer.  While this seems overly restrictive,
being able to answer the right yes/no question typically enables one to
also find an actual solution to the problem. (E.g., one can ask whether
the solution is in a certain interval, and divide this interval by two in
every step.) So let us get started with the complexity classes:
\begin{itemize}
\item 
\textsf{P} (``polynomial time''): The class of problems which can be
solved in a time which scales at most as some polynomial
$\mathrm{poly}(n)$ in the size of the problem.  For instance, multiplying
two numbers is in \textsf{P} -- the number of elementary operators scales
like the product of the number of digits of the two numbers to be
multiplied.  Generally, we consider the problems in \textsf P those which
are efficiently solvable.
\item \textsf{NP} (``non-deterministic polynomial time''): The class of all
decision problems where for ``yes'' instances, there exists a
``proof'' (i.e., the solution to the problem) whose correctness can be
\emph{verified} in a time which is $\mathrm{poly}(n)$, while for ``no''
instances, no such proof exists.  Typical problems are decomposing a
number into its prime factors (given the prime factors, their product can
easily be verified), or checking whether a given graph can be colored with
a certain number of colors without coloring two adjacent vertices in the
same color (for a given coloring, it can be efficiently checked whether
all adjacent vertices have different colors).  
\item \textsf{BQP} (bounded-error quantum polynomial time), the quantum
version of \textsf{P}: The class of problems which can be
(probabilistically) solved by a quantum computer in time
$\mathrm{poly}(n)$; a famous example of a problem in \textsf{BQP} which is
not known to be in \textsf{P} is Shor's algorithm for
factoring~\cite{shor:shor_alg}.  
\item \textsf{QMA} (quantum Merlin Arthur), one quantum version of
\textsf{NP}: The class of decision problems where for a ``yes'' instance,
there exists a quantum proof (i.e., a quantum state) which can be
efficiently verified by a quantum computer, and no such proof exists for a
``no'' instance.  For example, determining whether the ground state
energy of a local Hamiltonian $H$ is below some threshold $E$ (up to a
certain accuracy) is in \textsf{QMA} -- in a ``yes'' case, the proof would
be any state $\ket\psi$ with energy below the threshold, and the energy
$\bra\psi H \ket\psi$ can be efficiently estimated by a quantum computer,
using e.g.\ phase estimation~\cite{kitaev:book}.
\end{itemize}

\begin{figure}
\centering
\includegraphics[width=4cm]{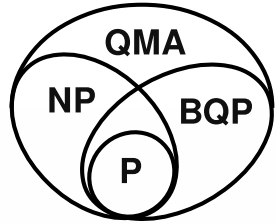}
\caption{\label{fig:cplxclasses}
Relevant complexity classes and their relation.}
\end{figure}

The relation of the classes to each other is illustrated in
Fig.~\ref{fig:cplxclasses}.  All of the inclusions shown are generally
believed to be strict, but none of them has up to now been rigorously
proven. In particular, whether \textsf{NP}$\ne$\textsf P is probably the
most famous open problem in theoretical computer science.

An important concept it that of complete problems: A problem is complete
for a class if it is as hard as any problem in the class. Completeness is
established by showing that any problem in the class can be \emph{reduced}
to the complete problem, this is, it can be reformulated as an instance of
the complete problem in $\mathrm{poly}(n)$ time. The relevance of complete
problems lies in the fact that they are the hardest problems in a class:
This is, if a problem is e.g.\ \textsf{NP}-complete, and we believe that
\textsf{NP}$\ne$\textsf P, then this problem cannot be solved in
polynomial time; and on similar grounds, \textsf{QMA}-complete problems
cannot be solved in polynomial time by either a classical or a quantum
computer. We will discuss physically relevant complete problems relating
to quantum many-body systems in the next section.

\subsection{The complexity of problems in many-body physics}

\begin{figure}[t]
\centering
\includegraphics[height=2.3cm]{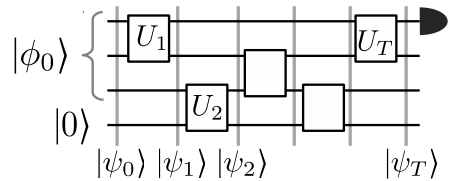}
\caption{\label{fig:qma-verifier}
QMA verifier: The proof $\ket{\phi_0}$, together with some ancillas
$\ket{0}$, is processed by the verifying circuit $U_T\cdots U_1$; the
intermediate state at step $t$ is denoted by $\ket{\psi_t}$.  At $t=T$,
the first qubit is measured in the computational basis to determine
whether the proof is accepted.
}
\end{figure}

In the following, we will discuss the computational complexity of
quantum many-body problems. The main problem we will consider is
determining whether the ground state energy of a Hamiltonian $H$ which is
a sum of few-body terms is below a certain threshold
(up to a certain accuracy).  As we have argued
above, this problem is inside the class \textsf{QMA}.  In the following,
we will argue that the problem is in fact \textsf{QMA}-complete, which
implies that it is hard not only for classical, but even for quantum
computers; the following construction is due to
Kitaev~\cite{kitaev:book,kitaev:qma}. To this end, we need to consider an
arbitrary problem in \textsf{QMA} and show that we can rephrase it as
computing the ground state energy of a certain Hamiltonian.  A problem in
\textsf{QMA} is described by specifying the quantum 
circuit which verifies the
proof, and asking whether it has an input which will be accepted with a
certain probability.  The verifying circuit is illustrated in
Fig.~\ref{fig:qma-verifier}.  Now let us assume that there is a valid proof
$\ket{\phi_0}$, and run the verifier with $\ket{\phi_0}$ (plus ancillas
$\ket0$) as an input. At each time step $t$, the state of the register the
verifier acts on is denoted by $\ket{\psi_t}$, and the proof is accepted
if the first qubit of the state $\ket{\psi_T}$ at time $T$ is in state
$\ket{1}$.

Our aim will now be to build a Hamiltonian which has the ``time history
state'' 
\begin{equation}
\label{eq:timehist-state}
\ket\Psi = \sum_t \ket{\psi_t}_R\ket{t}_T
\end{equation}
as a ground state. To
this end, we will use three types of terms: One term,
$H_\mathrm{init}$, will ensure that the ancillas are in the state
$\ket{0}$ at time $t=0$, by increasing the energy of non-zero ancillas if
$t=0$. A second term, $H_\mathrm{prop}$, will ensure correct propagation
from time $t$ to $t+1$, i.e., $\ket{\psi_t}=U_t\ket{\psi_{t-1}}$. Finally,
a last term $H_\mathrm{final}$ will increase the energy if the proof is
not accepted, i.e., the first bit of $\ket{\psi_T}$ is not $\ket1$.   Together, 
\begin{equation}
\label{eq:qma-ham}
H=H_\mathrm{init}+H_\mathrm{prop}+H_\mathrm{final}
\end{equation}
has a ground state with minimal energy if there exists a valid proof
[which in that case is exactly the time history state $\ket\Psi$ of
Eq.~(\ref{eq:timehist-state})].  On the other hand, in a ``no'' instance
in which no valid proof exists, any ground state -- which is still of the form
(\ref{eq:timehist-state}) -- must either start with incorrectly initialized
ancillas, or propagate the states incorrectly (violating 
 $\ket{\psi_t}=U_t\ket{\psi_{t-1}}$), or be rejected by
$H_\mathrm{final}$, thereby increasing the energy sufficiently to reliably
distinguish ``yes'' from ``no'' instances.  Together, this demonstrates that
finding the ground state energy of the Hamiltonian (\ref{eq:qma-ham}) is a
\textsf{QMA}-complete problem, showing that it is impossible to solve this
problem even for a quantum computer~\cite{kitaev:book,kitaev:qma}).

While the Hamiltonian (\ref{eq:qma-ham}) is not spatially local, there has
been a number of works sequentially proving \textsf{QMA}-hardness for more
and more simple Hamiltonians.  In particular, nearest-neighbor
Hamiltonians on a 2D square lattice of qubits are
\textsf{QMA}-complete~\cite{terhal:lh-2d-qma}, and, most notably, nearest
neighbor Hamiltonians on one-dimensional chains~\cite{aharonov:1d-qma},
thus substantiating the difficulties encountered in searching for 
convergence proofs for simulation methods even in 1D.

Let us note that there is also a classical version of the above
construction, showing \textsf{NP}-com\-plete\-ness of \emph{classical} ground
state problems: In that case, we can use a ``history state''  
\[
\ket{\psi_0}\otimes \ket{\psi_1}\otimes \cdots \otimes \ket{\psi_T}
\]
with no coherence between different times, where we think of the states as
being arranged in columns.  The Hamiltonian is similar to the one before,
Eq.~(\ref{eq:qma-ham}), just that the individual terms are no more
conditioned on the value of a ``time'' register, but simply act on the
columns of the history state corresponding to the right time $t$.
(Sloppily speaking, the reason that we can do so is that all terms in the
Hamiltonian act in a classical basis, and classical information can be
cloned.) The resulting Hamiltonian is at most four-local, acting across
$2\times 2$ plaquettes of adjacent spins. Again, it has been shown using
differerent techniques that even classical two-dimensional models with
Ising-type couplings and local fields are still
\textsf{NP}-complete~\cite{barahona:spinglass-np}.

Given all these hardness results for determining ground state properties
of classical and quantum spin systems, one might wonder why in practice,
simulation methods often work very well.  First, let us note that these
hardness result tell us why in many cases, we cannot \emph{prove} that
numerical methods work: This would require to identify the
necessary conditions such as to exclude all hard instances, while still
keeping the relevant problems, and to find a proof relying exactly on
these conditions.  Indeed, there are many properties which
separate practical problems from the hard instances we discussed: For
instance, in practice one usually considers translationally invariant (or
nearly translationally invariant) systems (though there exist hardness
results for translationally invariant systems even in one
dimension~\cite{gottesman:tinv-1d-qma-exp}), and the systems considered will
have various symmetries.   Even more importantly, all the
\textsf{QMA}-complete models known have a spectral gap of at most
$1/\mathrm{poly(n)}$, as opposed to the typically constant gap in many
problems of interest.  On the other hand, the \textsf{NP}-complete Hamiltonian
outlined above is completely classical and thus has gap $1$, while still
being \textsf{NP}-complete.  Overall, while there is no doubt that
numerical methods are extremely successful in practice, computational
complexity tells us that we have to carefully assess their applicability
in every new scenario. It should also be pointed out that the the fact
that a family of problems (such as ground state problems) is complete for
a certain complexity class addresses only the worst-case complexity, and
in many cases, typical instances of the same problem can be solved much
more efficiently.

\section{Summary}

In this lecture, we have given an overview over condensed matter
applications of quantum information methods, and in particular of
entanglement theory. The main focus was on entanglement-based ansatzes for
the description and simulation of quantum many-body systems.  We started
by discussing the area law for the entanglement entropy which is obeyed by
ground states of local interactions, and used this to derive the Matrix
Product State (MPS) ansatz which is well suited to describe the physics of
such systems.  We showed that the one-dimensional structure of MPS allows
for the efficient evaluation of expectation values, and that this can be
used to build a variational algorithm for the simulation of
one-dimensional systems.  We have also shown how MPS appear as 
ground state of local ``parent Hamiltonians'' which makes them suitable to
construct solvable models, and thus for the study of specific models as
well as the general structure of quantum phases. 

We have then discussed Projected Entangled Pair States (PEPS), which
naturally generalize MPS and are well suited for the description of
two-dimensional systems, and we have shown how approximation methods can
be used to implement efficient PEPS based simulations.  Again, PEPS can
also be used to construct solvable models very much the same way as for
MPS. We have subsequently demonstrated that MPS and PEPS can be used to
simulate the time evolution and thermal states of systems governed by
local Hamiltonians.  Moreover, we have discussed other tensor network
based approaches, such as MERA for scale-invariant systems or iMPS and
iPEPS for infinite sytems, and concluded with a discussion on how to apply
tensor network methods to fermionic systems.

Finally, we have introduced another field in which quantum information
tools are useful to understand condensed matter systems, namely the field
of computational complexity. Complexity tools allow us to understand which
problems we can or cannot solve. In particular, we have discussed the
class \textsf{QMA} which contains the problems which are hard even for
quantum computers, and we have argued that many ground
state problems, including 2D lattices of qubits and 1D chains, are
complete for this class.  While this does not invalidate the applicability
of simulation methods to these problems, it tells us that special care has
to be taken, and points out to us why we cannot hope for convergence
proofs without imposing further conditions.


\end{document}